\documentclass[10pt,twocolumn,letterpaper]{article}

\usepackage{iccv}
\usepackage{times}
\usepackage{epsfig}
\usepackage{graphicx}
\usepackage{amsmath}
\usepackage{amssymb}
\usepackage{mydefs}
\usepackage{rotating}

\usepackage[breaklinks=true,bookmarks=false]{hyperref}

\iccvfinalcopy 

\def\iccvPaperID{63} 

\ificcvfinal\fi

\begin{document}

\title{Frequency Separation for Real-World Super-Resolution}
\newcommand{\aand}{\hspace{6mm}}
\author{Manuel Fritsche \aand Shuhang Gu \aand Radu Timofte\\
		Computer Vision Lab, ETH Z\"urich, Switzerland \\
		{\tt\small \{manuelf, shuhang.gu, radu.timofte\}@ethz.ch}
	}

\maketitle
\ificcvfinal\thispagestyle{empty}\fi

\begin{abstract}
    Most of the recent literature on image super-resolution (SR) assumes the availability of training data in the form of paired low resolution (LR) and high resolution (HR) images or the knowledge of the downgrading operator (usually bicubic downscaling).
    While the proposed methods perform well on standard benchmarks, they often fail to produce convincing results in real-world settings. This is because real-world images can be subject to corruptions such as sensor noise, which are severely altered by bicubic downscaling. Therefore, the models never see a real-world image during training, which limits their generalization capabilities. Moreover, it is cumbersome to collect paired LR and HR images in the same source domain.

    To address this problem, we propose DSGAN to introduce natural image characteristics in bicubically downscaled images. It can be trained in an unsupervised fashion on HR images, thereby generating LR images with the same characteristics as the original images. We then use the generated data to train a SR model, which greatly improves its performance on real-world images. Furthermore, we propose to separate the low and high image frequencies and treat them differently during training. Since the low frequencies are preserved by downsampling operations, we only require adversarial training to modify the high frequencies. This idea is applied to our DSGAN model as well as the SR model. We demonstrate the effectiveness of our method in several experiments through quantitative and qualitative analysis. Our solution is the winner of the AIM Challenge on Real World SR at ICCV 2019.
\end{abstract}

\section{Introduction}
\label{sec:introduction}
\begin{figure}[!t]
    \centering  
	\begin{minipage}[c]{0.3\linewidth}
	    \includegraphics[width=\linewidth]{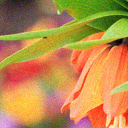}
	    \vspace{-0.04cm}
	    \centering \footnotesize original
	    \vspace{0.2cm}
	\end{minipage}
	\begin{minipage}[c]{0.3\linewidth}
	    \includegraphics[width=\linewidth]{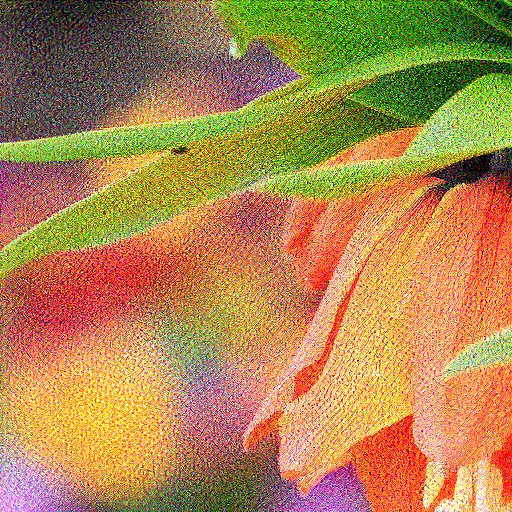}
	    \centering \footnotesize ESRGAN
	    \vspace{0.2cm}
	\end{minipage}
	\begin{minipage}[c]{0.3\linewidth}
	    \includegraphics[width=\linewidth]{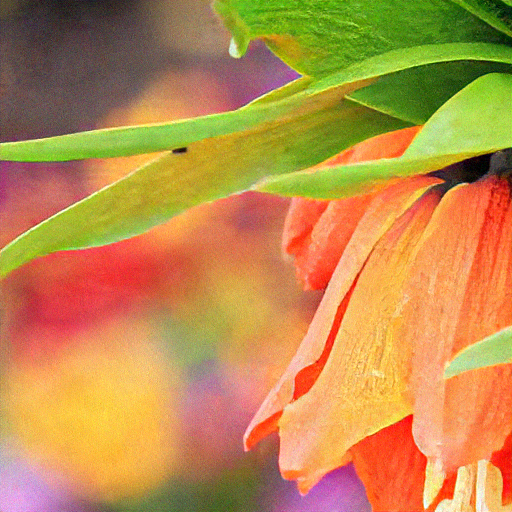}
	    \centering \footnotesize ours
	    \vspace{0.2cm}
	\end{minipage}
    \caption{$\times4$ SR comparison of ESRGAN~\cite{WangYWGLDQL18} and our method applied on a noisy input image. ESRGAN amplifies the corruptions, while our model preserves the noise level in the output.}
    \label{fig:esrgan_comparison_clean_corrupted}    
    \vspace{-0.5cm}
\end{figure}

The goal of image super-resolution (SR) is to increase the resolution in images. With the advent of convolutional neural networks (CNNs), the field has received increasing attention over the last couple of years. Modern techniques are now able to generate photo-realistic results on clean benchmark datasets. However, most state-of-the-art models~\cite{zhang2019ranksrgan, WangYWGLDQL18, LimSKNL17} perform poorly on real-world images, which can be subject to corruptions such as sensor-noise. These characteristics usually lead to strange artifacts in the super-resolved images as shown in Figure~\ref{fig:esrgan_comparison_clean_corrupted}.

The reason for this lies in the way these SR models are trained. Most of them rely on supervised training, which requires high resolution (HR) and corresponding low resolution (LR) image pairs. Since it is difficult to collect HR and LR images of the exact same scene, LR images are typically generated from HR images. In most cases, this is done by simply applying bicubic downscaling on HR images. While this method is easy and provides good results in clean settings, it comes with a significant problem: bicubic downscaling alters image characteristics. For example, it reduces corruptions in LR images, \ie makes them ``cleaner". Therefore, the model is only trained for input LR images that are altered by the downsampling operator. 
This leads to a significant performance drop when the model is applied on images that are not bicubically downscaled. 

Since many real-world images have visible corruptions, the state-of-the-art SR methods are not very useful in practice. Current generations of smartphones are equipped with hardware that allows the deployment of powerful neural networks. Therefore, robust SR methods can be very useful for improving the quality of images taken from smartphone cameras. This work is focused on improving the performance of SR models in such real-world settings. 

To achieve this, we aim to generate LR images that have the same characteristics as the images we want to super-resolve. These images allow SR models to be trained with a similar type of data that they encounter during application. In the first step, we create LR images by downsampling the HR images with bicubic downscaling. In a second step, we alter the characteristics of these LR images to match those of the source images. This is done by using a Generative Adversarial Network (GAN) setup~\cite{goodfellow2014generative}, which allows us to train a neural network to make our LR images indistinguishable from the source images. However, training a GAN is very difficult and needs to be stabilized to converge to the desired result. Similar to Ignatov~\etal~\cite{ignatov2017dslr,Ignatov_2018_CVPR_Workshops}, we achieve this stabilization by combining multiple loss functions: a color loss that forces the network to keep the low frequencies of the original image and an adversarial loss that uses the discriminator to produce high frequencies that look similar to the ones in the source images. Finally, we also add a perceptual loss that pushes the output towards solutions that look similar to the input images. Thereby, it ensures that the high frequencies generated by the GAN are still matching the low frequencies that are supervised by the color loss. This setup is based on the following idea:
during the process of downsampling an image, the high frequencies are removed, while the low frequencies remain. Therefore, the resulting LR images lack the high frequency characteristics found in the original images. On the other hand, low image frequencies such as the color are preserved to an extend that depends on the downscaling factor. 
By limiting the adversarial loss to high frequencies, we greatly reduce the complexity of the task. This helps the discriminator to focus on the relevant image features and leaves others untouched. Therefore, our setup is more stable, converges faster and produces better results than a standard GAN.

Furthermore, we also apply our idea of separating the low and high image frequencies to train the SR model. Thereby, we use a similar strategy as mentioned above: use a pixel-wise loss to stabilize the low frequencies and apply the adversarial loss only on the high frequencies. Since this separates the pixel-wise and the adversarial loss, it simplifies the task of the discriminator. We also provide theoretical reasoning for why it makes sense to only train the high frequencies with a GAN and use a simple pixel-wise loss for the low frequencies.

We evaluate our methods on multiple datasets with artificial and natural corruptions\footnote{Our code and models are publicly available at \url{https://github.com/ManuelFritsche/real-world-sr}}. To show the effectiveness of our implementation, we use the DF2K dataset that is a combination of the DIV2K~\cite{agustsson2017ntire,timofte2017ntire} and Flickr2K~\cite{LimSKNL17} datasets. Since this dataset contains clean images, it allows us to introduce artificial corruptions and create ground truth (GT) HR and LR image pairs by adding the same corruptions to both of them. We ran experiments with sensor noise as well as compression artifacts. In both cases we demonstrate the effectiveness of our methods through quantitative and qualitative evaluations. Furthermore, we ran our methods on real-world images from the DPED dataset~\cite{ignatov2017dslr}, which were collected by an iPhone 3 camera. We only provide qualitative evaluation in this case since no GT is available. 
Finally, we also participated in the AIM 2019 Challenge on Real World Super-Resolution~\cite{AIM2019RWSRchallenge} associated with the AIM workshop at ICCV 2019. Our method won the first place in both tracks for source domain and target domain. 
None of our methods are specifically designed for a certain type of data. They can also be applied to images with other characteristics than the ones we used in our experiments.

\section{Related Work}
\label{sec:related_work}
In recent years, the field of image super-resolution has been dominated by CNNs, which achieve state-of-the-art performance~\cite{timofte2017ntire,Timofte_2018_CVPR_Workshops,cai2019ntire}. Dong~\etal~\cite{DongLHT14, DongLHT16} introduced the first top performing CNNs trained end-to-end to map LR to HR images. Based on this pioneering work, several refinements have been proposed~\cite{KimLL16, KimLL16a, TaiY017, LaiHA017}. Thereby, deeper networks with residual layers such as EDSR~\cite{LimSKNL17} produce better results than standard CNNs. Additional improvements were made by using different variants of densely connected residual blocks~\cite{zhang2018residual, WangYWGLDQL18} as building blocks of the model. These blocks allow to further increase the depth of the networks, resulting in very powerful models.

Most of the previously mentioned methods are based on optimizing the $L_1$ or $L_2$ distance between the SR image and the ground truth HR image. While this strategy achieves state-of-the-art performance in image fidelity metrics such as PSNR, the resulting images are often blurry. This is because the human perception of visual similarity only has a limited correlation with such pixel-wise errors. Therefore, more recent SR methods are based on loss functions and training methods that are better suited to produce visually pleasing images. Gatys~\etal~\cite{gatys2015texture, gatys2016image} show that high-level features extracted from pre-trained networks can be used to design perceptual loss functions. Such a loss function is used in~\cite{JohnsonAF16} to enhance the visual quality of super-resolved images. The SRGAN model~\cite{LedigTHCCAATTWS17} is trained with an additional adversarial loss to push the output to the manifold of natural images, which allows to generate photo-realistic results. Several works have proposed further improvements with approaches that focus on perceptual similarity~\cite{SajjadiSH17, WangYWGLDQL18, zhang2019ranksrgan}. The recently introduced RankSRGAN~\cite{zhang2019ranksrgan} uses a method to train SR models on indifferentiable perceptual metrics. Our experiments are based on ESRGAN~\cite{WangYWGLDQL18}, the winner of the PIRM 2018 challenge on perceptual super-resolution~\cite{Blau_2018_ECCV_Workshops}. It introduces several improvements to the SRGAN model, thereby achieving state-of-the-art perceptual performance. 

All of the previously mentioned models are trained with HR/LR image pairs, generated through bicubic downscaling. Therefore, these models perform poorly in real-world scenarios. One way of addressing this issue is by directly collecting paired data, which is explored in recent work~\cite{cai2019ntire, chen2019camera}. However, these approaches rely on complicated hardware and require the collection of new data for each camera source.
Other methods try to make SR more robust, tailor them to the test image. Liang~\etal~\cite{LiangTWGZ17} proposed to fine-tune a pre-trained SR model to the test image. 
ZSSR~\cite{ShocherCI18} is a lightweight CNN that is trained by only using the test image, which allows the network to focus only on image specific details. However, both approaches still rely on a known downsampling operation during training. 
Additionally, training a network for each test image results in very slow predictions. Yuan~\etal~\cite{YuanLZZDL18} propose a model that learns a mapping from the original input to a clean input space, after which they apply super-resolution. They use a complex framework with two cycle-consistency losses, which increases training time. Their initial cleaning step improves the performance of the model on corrupted images, but also increases the complexity of their model. Conversely, our approach mainly focuses on generating training data. We only make small modifications in the discriminator and loss functions, which does not introduce more complexity in the model. 
Similar to our work, some novel methods generate paired data artificially.
Kim~\etal~\cite{KimCLL18} propose an auto-encoder-based framework to jointly learn downsampling and upsampling. While their super-resolution method performs well on images downsampled by their model, it is not applicable to unknown downsampling operations.
Bulat~\etal~\cite{BulatYT18} explore a method to learn the downsampling operation. However, they focus only on faces, but not the general super-resolution problem, which makes the task a lot easier. In contrast, we do not make any assumptions on the content of the images.

\section{Proposed Method}
\label{sec:proposed_method}

\subsection{Real-World Super-Resolution}
\label{sec:real_world_sr}
\begin{figure}[!t]
    \centering
    \begin{minipage}[c]{0.3\linewidth}
	    \includegraphics[width=\linewidth]{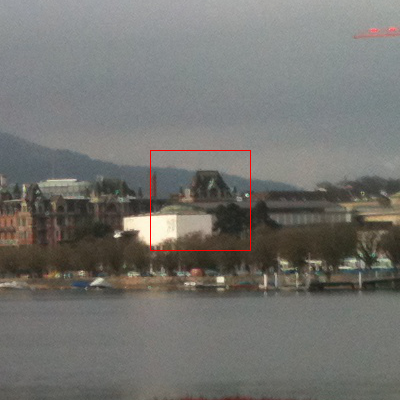}
	    \centering \footnotesize original
	    \vspace{0.2cm}
	\end{minipage}
    \begin{minipage}[c]{0.3\linewidth}
	    \includegraphics[width=\linewidth]{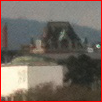}
	    \centering \footnotesize cropped
	    \vspace{0.2cm}
	\end{minipage}
	\begin{minipage}[c]{0.3\linewidth}
	    \includegraphics[width=\linewidth]{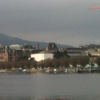}
	    \centering \footnotesize bicubic
	    \vspace{0.04cm}
	    \vspace{0.2cm}
	\end{minipage}
    \caption{This image is taken from the DPED dataset~\cite{ignatov2017dslr} and shows strong corruptions. When comparing a cropped region (middle) to the bicubically downscaled version of equal size (right), one can clearly see how this operation reduces corruptions.}
    \label{fig:real_world_bicubic}    
\end{figure}

State-of-the-art SR models rely on fully supervised training of neural networks on paired HR and LR images. While collecting images is not a difficult task, obtaining paired images from two different sources is difficult and cumbersome. For this reason, the SR field mainly relies on using bicubic downscaling to generate image pairs. Although this approach has helped the development of promising SR models, it is limiting the generalization to real-world images because it can drastically alter certain image characteristics such as sensor noise. An example of how this downsampling operation affects a real-world image can be seen in Figure~\ref{fig:real_world_bicubic}.

For our analysis, we assume that we are given a set of \textit{source} images, which have similar characteristics (\eg the same sensor noise distribution). One can then define the HR images $\y\in\Y$ either directly as the source images or as modified versions thereof. Finally, the LR images $\x\in\X$ are generated by downsampling the HR images. As traditional downsampling operations alter the image characteristics, images from $\X$ differ from images in $\Z$. Since our goal is to upsample images from the domain $\Z$, we aim to have $\X=\Z$. In other words, the LR images seen in training should be indistinguishable from the source images. Therefore, we aim to map images from $\X$ to $\Z$, while preserving the image content.

\subsection{Downsampling with Domain Translation}
\label{sec:downscaling_with_domain_translation}
\begin{figure}[!t]
    \centering  
    \includegraphics[width=\linewidth]{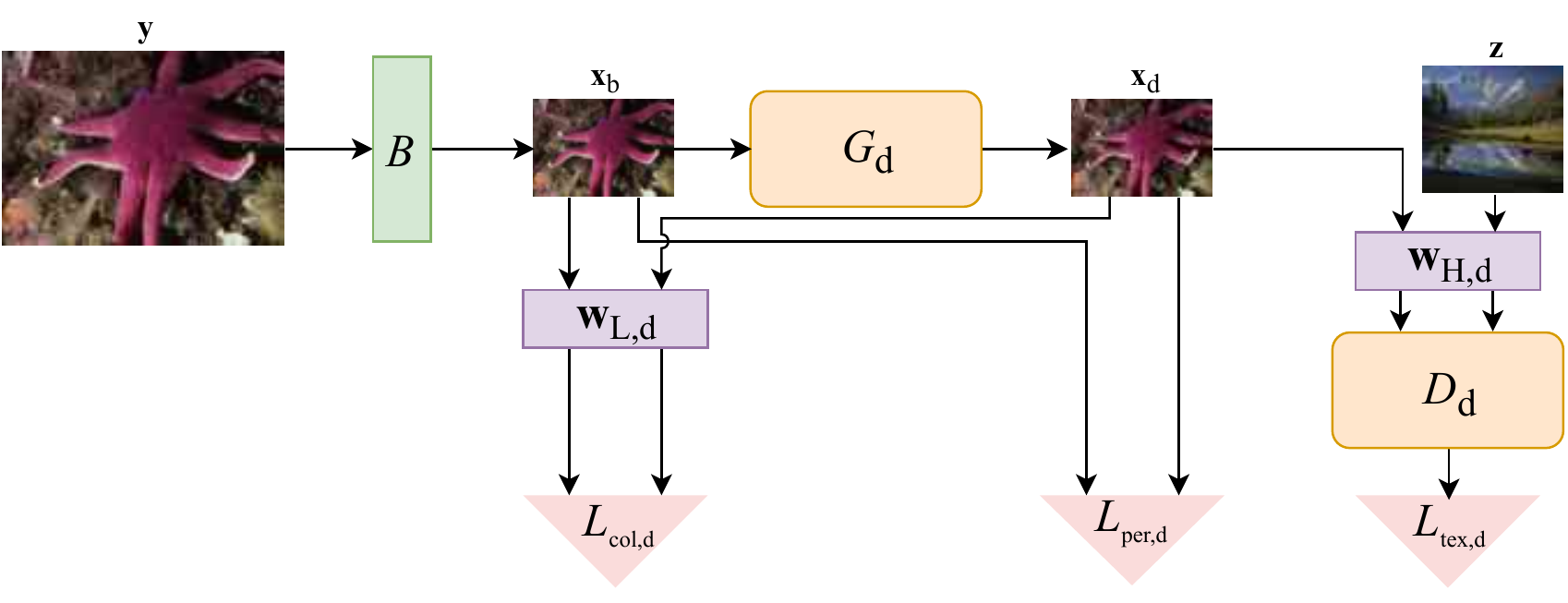} 
    \caption[Downsampling Structure]{Visualizes the structure of the downsampling setup. $B$ denotes the bicubic downscaling method, while the purple fields display the high- and low-pass filters. The red triangles denote the loss functions and the orange fields the neural networks.}  
    \label{fig:downsampling_structure}    
\end{figure}

In the following, we describe a model that can produce realistic LR images in the source domain $\Z$, given HR images in some $\Y$ domain. The complete overall structure is shown in Figure~\ref{fig:downsampling_structure}.

In the first step, we bicubically downscale the HR image $\y$ to obtain a LR image $\xb = B(\y)$. Since $\xb$ is now in the wrong domain $\X$, we use a neural network $\gdd$ to translate it to the source domain $\Z$. We call this new image $\xd$ with $\xd=\gend{\xb}$. To train $\gdd$ we use a standard GAN~\cite{goodfellow2014generative} setup with an additional discriminator $\ddd$. The discriminator is trained with the generator output $\gend{B(\y)}$ as fake data and uses the source images $\z\in\Z$ as real data.

\paragraph{Network Architectures}
The generator network consists mainly out of residual blocks~\cite{he2016deep} with two convolutional layers and ReLU activations in between. Except for the output layer, all convolutional layers use a $3\times 3$ kernel with $64$ features.

As image characteristics do not change the global image content, but only introduce local changes, we use a patch-based discriminator~\cite{li2016precomputed, isola2017image}. This discriminator is fully convolutional and returns a 2D array that determines which regions of the image look real or fake. The discriminator applies four convolutional layers with $5\times 5$ kernels on the low-pass filtered input image. The number of output features of each convolutional layer increases from $64$ to $128$ and $256$ with the final layer only using one feature map. Between the convolutional layers we apply Batch Normalization and LeakyReLU activations.

\subsection{Frequency Separation}
\label{sec:downscaling_frequency_separation}
As described in Section~\ref{sec:downscaling_with_domain_translation}, we are using a standard GAN setting to translate the original LR images to the source domain $\Z$. However, we do not just want any image in this source domain, but the one that is closest to our original LR image $\xb$. One way of achieving this is by using multiple loss functions. By introducing a perceptual and a pixel loss, one can restrict the possible solutions that the generator produces. Unfortunately, such a loss function is hard to balance because we need the output of the generator $\xd$ to stay close to the input $\xb$ and at the same time introduce the image characteristics of the source domain. The result is that we have to deal with a trade-off and neither of the goals will be achieved perfectly. 

This naive 
approach ignores the fact that we are dealing with downsampled images. As we discuss in Section~\ref{sec:frequency_separation}, the downsampling process removes the high image frequencies and keeps the low frequency information within a reduced number of pixels. This leads to high frequency characteristics being lost, while low frequency information such as color and context remain. As the low image frequencies are preserved, we only need to alter the high frequencies in our mapping from $\X$ to $\Z$. Therefore, we propose to apply the discriminator only on the high frequencies $\xhd$ of $\xd$ and keep the low frequencies $\xld$ close to the original ones. This greatly reduces the complexity of the problem, making it easier for the discriminator to focus on the relevant image features. In our experiments, we found this method to be crucial in training the GAN. It not only speeds up the training process, but also produces better results. Furthermore, our GAN setup reduces undesired color shifts, because the discriminator ignores the low image frequencies.

We separate the low and high image frequencies by using simple linear filters. For this purpose, we first define a low-pass filter $\wld$. The low and high frequencies can be obtained by simple convolutions:
\begin{equation}
    \label{eq:dsgan_low}
    \xld = \wld * \xd,
\end{equation}
\begin{equation}
    \label{eq:dsgan_high}
    \xhd = \xd - \xld = (\delta - \wld) * \xd.
\end{equation}
Therefore, we define our high-pass filter as $\whd=\delta - \wld$. After applying a high-pass filter, we feed the remaining frequencies $\xhd$ to the discriminator $\ddd$. The same high-pass filter is applied to the source images $\z\in\Z$. 
In our experiments, we empirically chose a moving average with kernel size $5$ as low-pass filter. However, our method is not limited to any specific filter. 

\paragraph{Loss Functions}
\label{sec:loss_functions}
We combine multiple loss functions to train our model. The generator loss combines three different losses: color loss $L_{\text{col,d}}$, perceptual loss $L_{\text{per,d}}$ and texture loss $L_{\text{tex,d}}$, which is an adversarial loss. 

The color loss is focusing on the low frequencies of the image. Since we do not want to change the low frequencies of $\xb$, we apply an $L_1$ loss on these frequencies to keep them close to the input. The color loss is defined as
\begin{equation}
    L_{\text{col,d}} = \frac{1}{m}\sum_{i=1}^m \left\lVert \wld * \gend{\xb^{(i)}}-\wld * \xb^{(i)}\right\rVert_1,
\end{equation}
where $m$ denotes the batch size.

As discussed in Section~\ref{sec:downscaling_frequency_separation}, we only apply the GAN loss on the high frequencies of the output $\xd$, which results in the following loss for generator and discriminator:
\begin{equation}
    L_{\text{tex,d}} = \m\frac{1}{m}\sum_{i=1}^m \mean{\log \discd{\whd * \gend{\xb^{(i)}}}},
\end{equation}
\begin{equation}
    \begin{aligned}
        L_{D_\text{d}} = &  \m\frac{1}{m}\sum_{i=1}^m\mean{\log \discd{\whd * \z^{(i)}}} \\
        & + \mean{\log\left(1-\discd{\whd * \gend{\xb^{(i)}}}\right)}.
    \end{aligned}
\end{equation}
Since the discriminator $\ddd$ returns a 2D array of values and not just a single one, we take the mean over all these values in the loss function.

Finally, to ensure that the high and low frequencies fit together, we also apply a perceptual loss $L_{\text{per,d}}$ to $\xb$ and $\gend{\xb}$. For this loss we use LPIPS~\cite{zhang2018unreasonable}, which is based on the features of the VGG network~\cite{simonyan2014very}.

We define the complete loss functions of the generator as
\begin{equation}
    L_{G_\text{d}} = L_{\text{col,d}} + 0.005\cdot L_{\text{tex,d}} + 0.01 \cdot L_{\text{per,d}}.
\end{equation}

\subsection{Frequency Separation for Super-Resolution}
\label{sec:frequency_separation}
\begin{figure}[!t]
    \centering  
    \includegraphics[width=\linewidth]{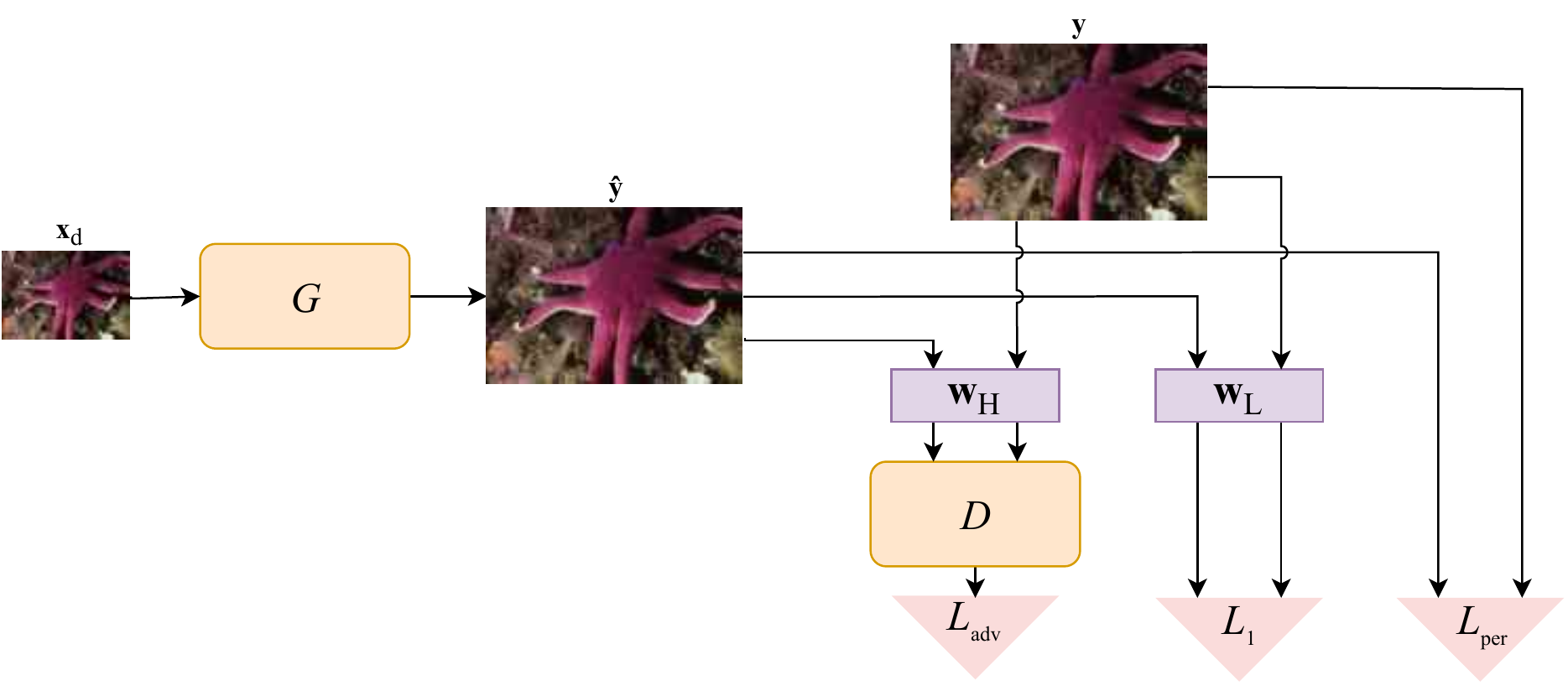} 
    \caption{Visualizes our changes to the ESRGAN structure. Only the purple blocks are added to filter the images.}  
    \label{fig:esrgan-fs_structure}    
\end{figure}

We also apply our idea of frequency separation directly on ESRGAN~\cite{WangYWGLDQL18}. However, the approach is not limited to this SR model and can easily be adapted for other methods. Our changes to the model are visualized in Figure~\ref{fig:esrgan-fs_structure}.

For our analysis, we look at an image $\y$ that we downsample by a factor $r$. Let us assume that $\x$ is a downsampled version of $\y$ without aliasing. 
The Sampling Theorem tells us that $\x$ allows us to infer the lowest $1/r$ fraction of the possible frequencies of the original image $\y$, which we denote as $\yl$ in the following. The remaining high frequencies are defined as $\yh=\y-\yl$.
There is no need to consider the context and generate fake details to map from $\x$ to $\yl$. In contrast to the mapping from $\x$ to $\y$, the mapping from $\x$ to $\yl$ is a one-to-one mapping, which allows it to be reconstructed directly. For reconstructing $\yh$ on the other hand, we can only rely on the context information, because it contains all the high frequencies that are removed by the downsampling and anti-aliasing process. Thus, the mapping from $\x$ to $\yl$ is considerably easier to learn than the mapping from $\x$ to $\yh$. 

Similar to Section~\ref{sec:downscaling_frequency_separation}, we use a low-pass filter $\wl$ and a high-pass filter $\wh$ to split up the low and high image frequencies. In our experiments, we found that a simple moving average with kernel size $9$ works well.
The low frequencies of $\gx$ can be learned directly over the $L_1$ loss, since there is only one possible $\yl$ given $\x$. Since the colors of an image are mainly defined in the low frequencies, we call this loss the color loss $L_{\text{col}}$:
\begin{equation}
    L_{\text{col}} = \frac{1}{m}\sum_{i=1}^m \left\lVert \wl * \gen{\x^{(i)}}-\wl * \y^{(i)}\right\rVert_1.
\end{equation}
The high frequencies of $\gx$ on the other hand have multiple ground truth values and cannot be learned through a pixel-based loss. Therefore, we use the adversarial loss only on these high frequencies, by simply adding a high-pass filter in front of the discriminator. This greatly reduces the complexity of the task, as the discriminator does not have to deal with the low frequencies. 

To make sure the high frequency details generated by the GAN loss match the low frequencies, the perceptual loss is applied on the full output. This results in the following adapted loss function for the ESRGAN generator:
\begin{equation}
\label{eq:mesrgan_complete_loss}
    L_{G}= L_{\text{per}} + 0.005 \cdot L_{\text{adv}} + 0.01 L_{\text{col}},
\end{equation}
This loss function simplifies the task of the discriminator, which allows the model to produce outputs that match the target distribution more closely.

\section{Experiments}
\label{sec:experiments}


\subsection{Experimental Setup}
\label{sec:experimental_setup}

\paragraph{Dataset Generation}
\label{sec:experimental_setup_dsgan}
For all experiments, we use a scaling factor of $4$ between the HR and LR images. We generate the HR/LR image pairs by using the model described in Section~\ref{sec:real_world_sr}, which we call \textit{DSGAN} (DownSampleGAN). We train it with $512 \times 512$ image patches, which we bicubically downscale with the MATLAB imresize method. For the discriminator, we use random $128 \times 128$ crops of the source images $\z\in\Z$. Using a batch size of $16$ image patches, we train the model for $200$ or $300$ epochs, depending on the size of the dataset. Similar to CycleGAN~\cite{ZhuPIE17}, we use Adam optimizer~\cite{kingma2014adam} with $\beta_1=0.5$ and an initial learning rate of $2\cdot 10^{\m 4}$. The learning rate is kept constant for the first half of the epochs and then linearly decayed to $0$ during the remaining epochs.

\textit{Same Domain Super-Resolution (SDSR):} In this setup, we aim to generate a training dataset with HR and LR images that are both in the source domain $\Y=\Z$. We use the source images directly as the HR images and train our DSGAN model to map from the domain of bicubically downscaled LR images $\X$ to the domain of the HR images $\Y$.

\textit{Target Domain Super-Resolution (TDSR):} If the images in the source domain have corruptions such as sensor noise, it is often desirable to remove these in the SR process. Therefore, in the TDSR setting, we aim to use HR images in a clean domain $\Y\neq\Z$ and only have our LR images in the source domain $\Z$. We generate the HR images by bicubically downscaling the source images with a factor of $2$. Since we use a scaling factor of $4$ between the HR and LR images, bicubic downscaling removes almost all corruptions. Therefore, we assume that the bicubically downscaled HR images from the SDSR setting and the TDSR setting to be approximately in the same domain $\X$. Thus, we apply the same model trained for SDSR in our TDSR setting to generate the LR images in the $\Z$ domain.

\paragraph{ESRGAN and ESRGAN-FS}
\label{sec:experimental_setup_esrgan}
In the second step, we use the HR/LR image pairs to train our SR model with either ESRGAN or our modified ESRGAN described in Section~\ref{sec:frequency_separation}, which we denote as \textit{ESRGAN-FS} in the following. 
We initialize training with the fully pre-trained ESRGAN weights in both cases, as training ESRGAN from scratch takes a long time. We then perform $50k$ training iterations with an initial learning rate of $10^{-4}$. The learning rate is halved after $5k$, $10k$, $20k$ and $30k$ iterations. We use Adam optimizer~\cite{kingma2014adam} with $\beta_1 = 0.9$ and $\beta_2=0.999$ for both generator and discriminator training.

\paragraph{Datasets}
\label{sec:experiments_datasets}
For our experiments with artificial corruptions, we use the DF2K dataset, which is a merger of the DIV2K~\cite{agustsson2017ntire,timofte2017ntire} and Flickr2K~\cite{LimSKNL17} datasets. It contains 3450 train images and 100 validation images with very little corruptions. 
To evaluate our method, we introduce two kinds of corruptions: sensor noise and compression artifacts. The sensor noise is modeled by adding pixel-wise independent Gaussian noise with zero mean and a standard deviation of $8$ pixels to the images. The compression artifacts are introduced by converting the images to JPEG with a quality of 30. Thereby, we create ground truth HR/LR pairs by applying the same degradation on both HR and LR images. 

To test our methods on real-world data, we use images from the DPED dataset~\cite{ignatov2017dslr}. More precisely, we use the $5614$ train and $113$ test images taken with an iPhone 3 camera. Since we cannot generate any ground truth for these images, we only compare the results visually.

Furthermore, we also participated in the AIM 2019 Challenge on Real World Super-Resolution~\cite{AIM2019RWSRchallenge}. In this challenge 2650 corrupted source images and 800 clean target images are provided for training. The corruptions in the source data are artificial but unknown. The validation and test set contain 100 images each and have the same corruptions as the source data~\cite{lugmayrICCVW2019}.

\paragraph{Quantitative Evaluation}
\label{sec:quantitative_evaluation}
For our quantitative evaluation, we use the popular PSNR and SSIM methods, which we calculate with the scikit-image measure module~\cite{van2014scikit}. While SSIM and PSNR are often used for measuring similarity to ground truth images, the resulting similarity values often correlate poorly with actual perceived similarity. Therefore, we also use LPIPS~\cite{zhang2018unreasonable} for comparison. As mentioned in Section~\ref{sec:loss_functions}, this measure is based on the features of pre-trained neural networks, which have been shown empirically to correlate better with human perception than handcrafted methods. We use the LPIPS method that is based on the features of AlexNet~\cite{krizhevsky2012imagenet} for evaluation.

\subsection{Comparison with State-of-the-Art}
\label{sec:comparison_with_sota}
In this section, we compare our methods with four other state-of-the-art methods. The first one is ESRGAN~\cite{WangYWGLDQL18}, where we report the results with and without additional fine-tuning on the corrupted dataset (using bicubic downscaling). The fine-tuned model is referred to as ESRGAN (FT). We also look at another method called RankSRGAN~\cite{zhang2019ranksrgan}. This method uses an additional model called Ranker to simulate the behavior of indifferentiable perceptual metrics. It then trains the model with these simulated perceptual metrics. We use the pre-trained weights based on the NIQE metric~\cite{mittal2012making}. Furthermore, we look at EDSR~\cite{LimSKNL17}, which is a method optimized for PSNR-based super-resolution. We also include ZSSR~\cite{ShocherCI18} in our comparison, which applies a Zero-Shot learning strategy on each image it super-resolves. 

\begin{figure}[!t]
    \centering  
	\begin{minipage}[c]{0.196\linewidth}
	    \includegraphics[trim=52px 0 52px 0, clip, width=\linewidth]{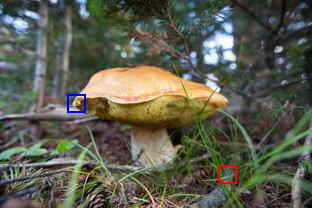}
	    \includegraphics[trim=52px 0 52px 0, clip, width=\linewidth]{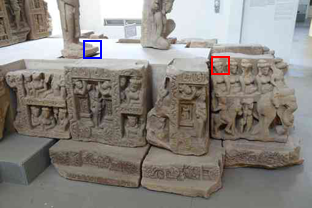}
	    \includegraphics[trim=34px 0 34px 0, clip, width=\linewidth]{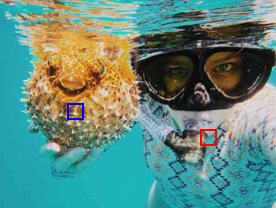}
	    \includegraphics[trim=34px 0 34px 0, clip, width=\linewidth]{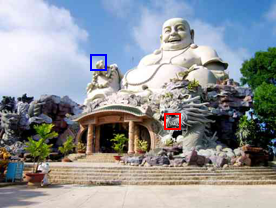}
	    \centering \tiny original \\ ~
	    \vspace{0.2cm}
	\end{minipage}
	\hspace{-0.02\linewidth}
	\begin{minipage}[c]{0.098\linewidth}
        \includegraphics[width=\linewidth]{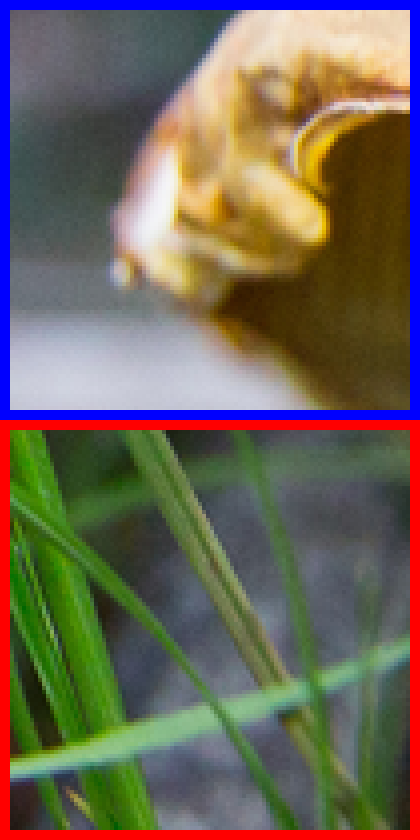}
        \includegraphics[width=\linewidth]{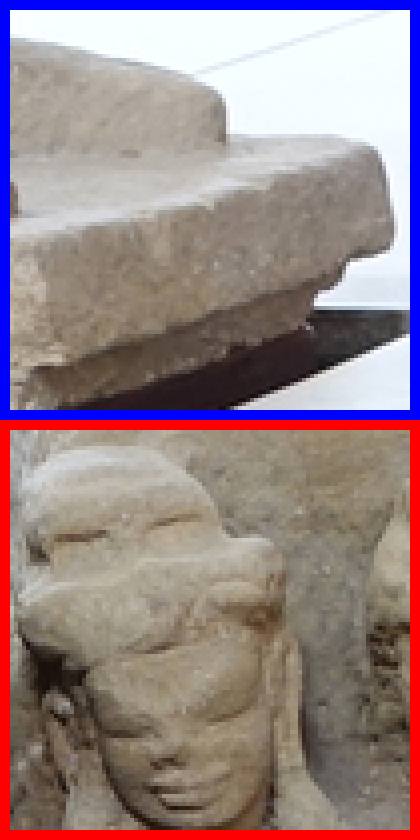}
        \includegraphics[width=\linewidth]{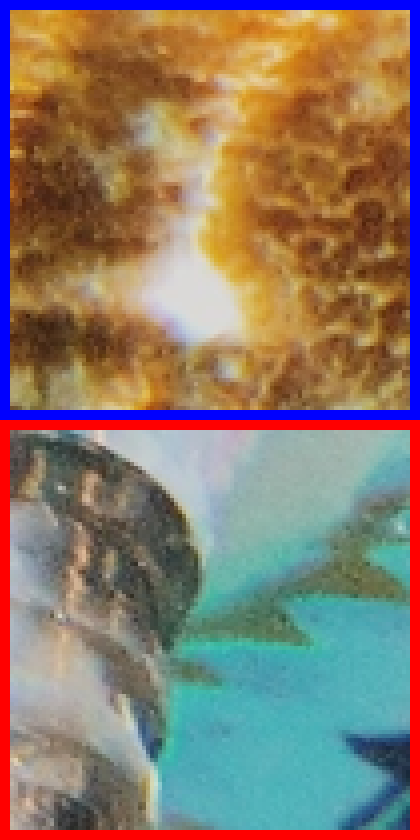}
        \includegraphics[width=\linewidth]{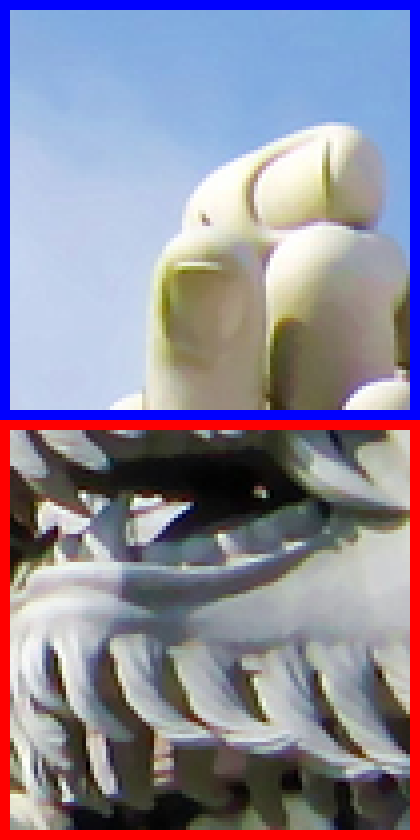}
        \centering \tiny GT \\ ~
        \vspace{0.2cm}
	\end{minipage}
	\hspace{-0.02\linewidth}
	\begin{minipage}[c]{0.098\linewidth}
        \includegraphics[width=\linewidth]{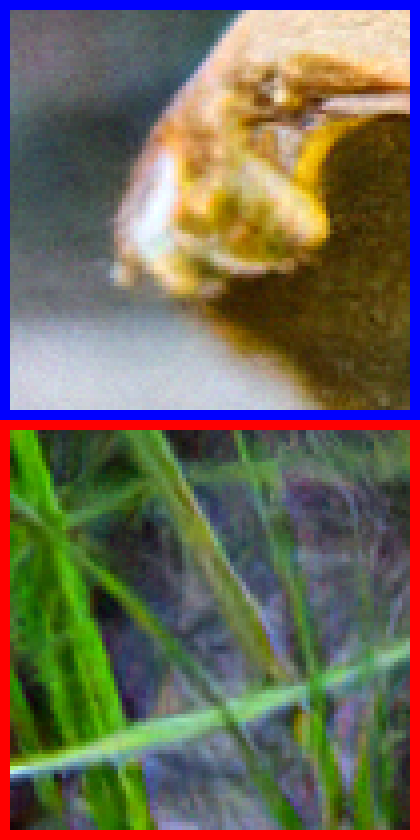}
        \includegraphics[width=\linewidth]{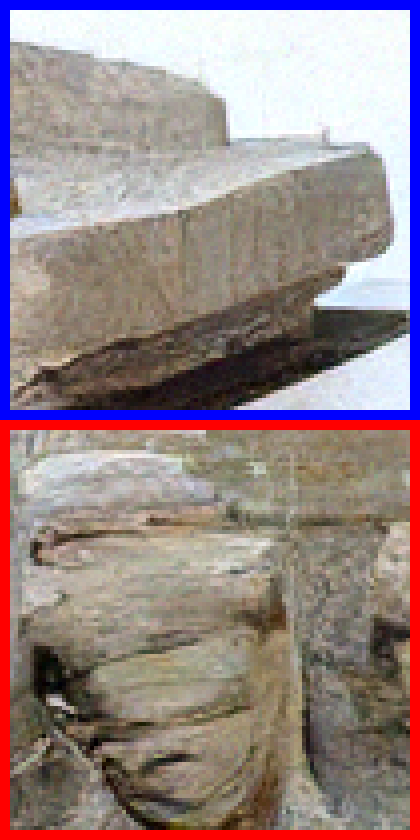}
        \includegraphics[width=\linewidth]{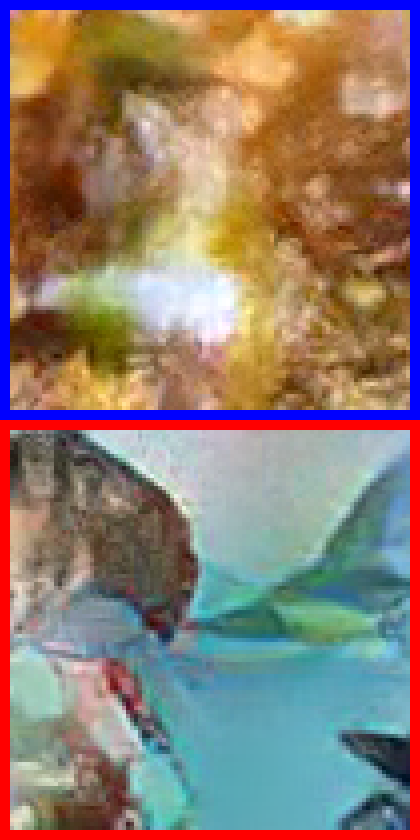}
        \includegraphics[width=\linewidth]{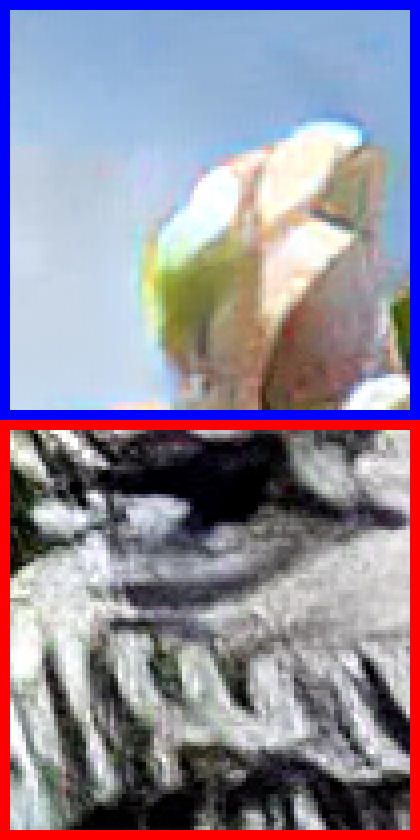}
        \vspace{-0.03cm}
        \centering \tiny \textbf{ours} \\ (TDSR)
        \vspace{0.2cm}
	\end{minipage}
	\hspace{-0.02\linewidth}
	\begin{minipage}[c]{0.098\linewidth}
        \includegraphics[width=\linewidth]{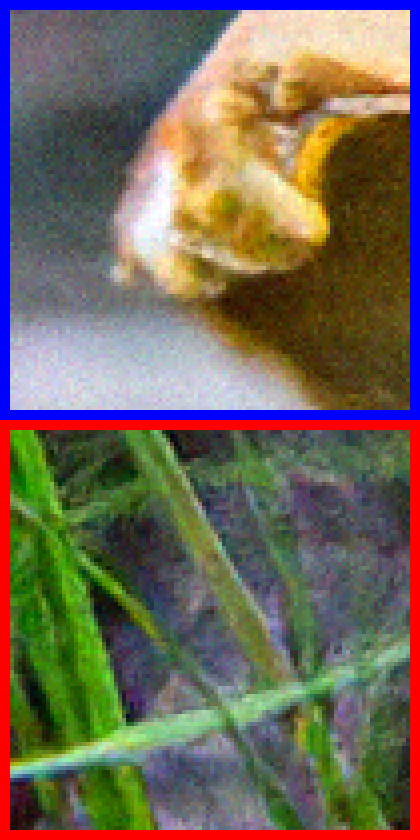}
        \includegraphics[width=\linewidth]{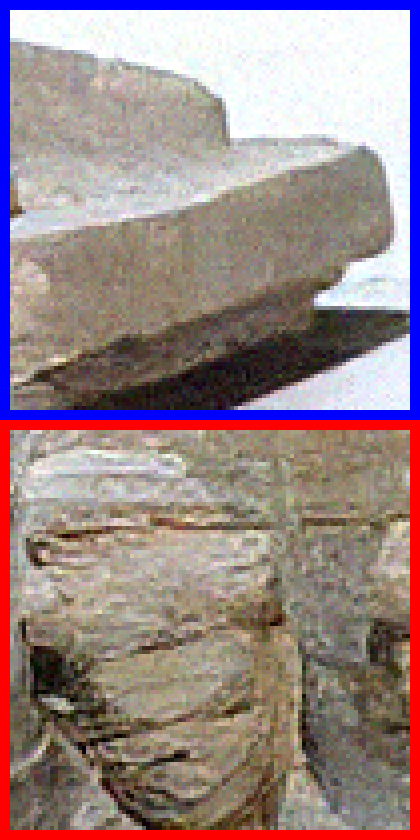}
        \includegraphics[width=\linewidth]{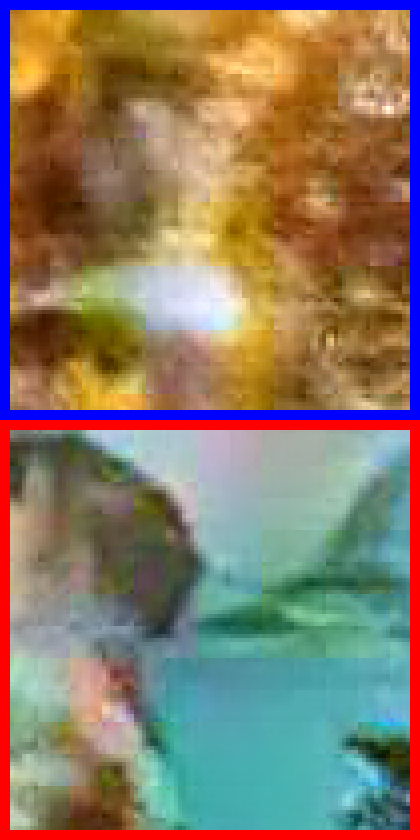}
        \includegraphics[width=\linewidth]{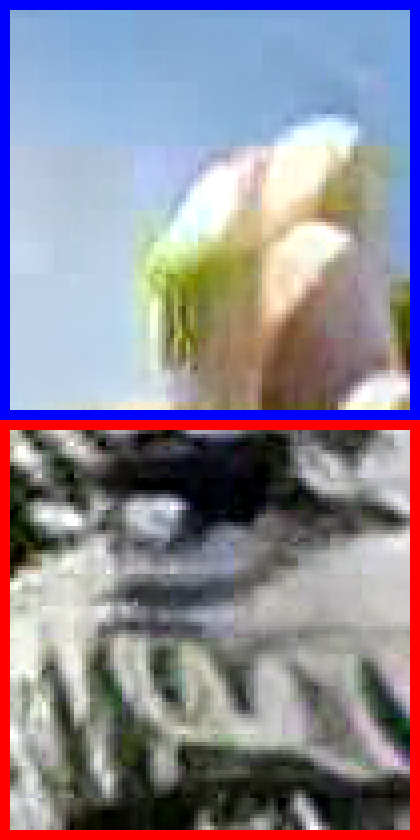}
        \vspace{-0.03cm}
        \centering \tiny \textbf{ours} \\ (SDSR)
        \vspace{0.2cm}
	\end{minipage}
	\hspace{-0.02\linewidth}
	\begin{minipage}[c]{0.098\linewidth}
        \includegraphics[width=\linewidth]{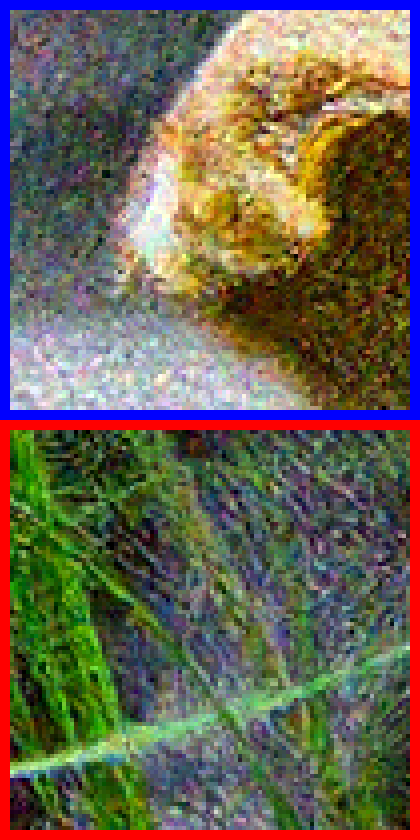}
        \includegraphics[width=\linewidth]{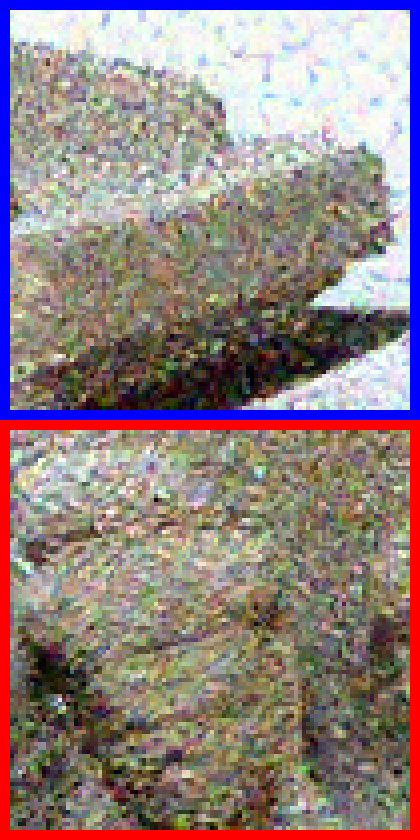}
        \includegraphics[width=\linewidth]{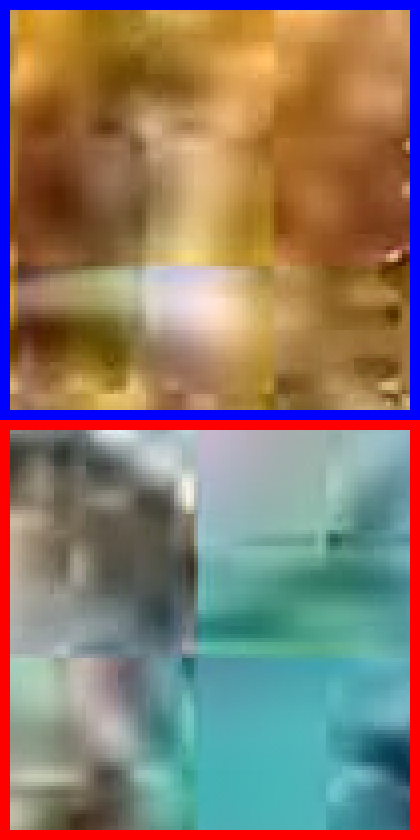}
        \includegraphics[width=\linewidth]{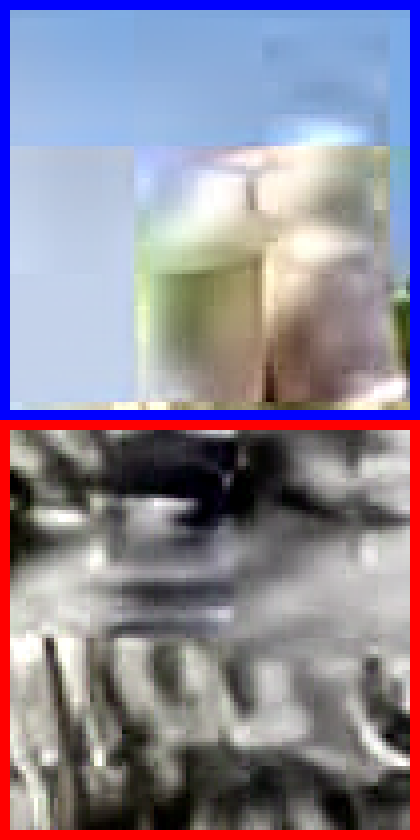}
        \vspace{-0.03cm}
        \centering \tiny ESRGAN \\ (FT)
        \vspace{0.2cm}
	\end{minipage}
	\hspace{-0.02\linewidth}
	\begin{minipage}[c]{0.098\linewidth}
        \includegraphics[width=\linewidth]{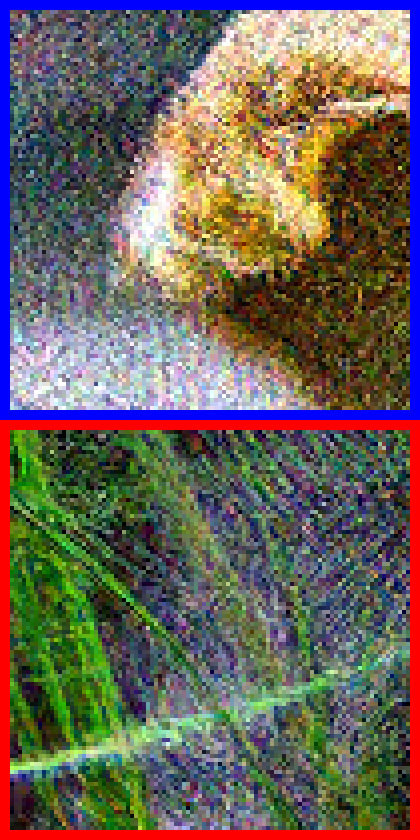}
        \includegraphics[width=\linewidth]{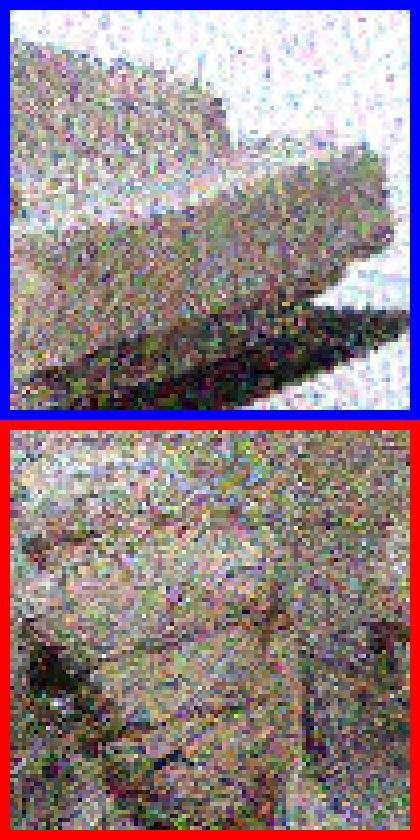}
        \includegraphics[width=\linewidth]{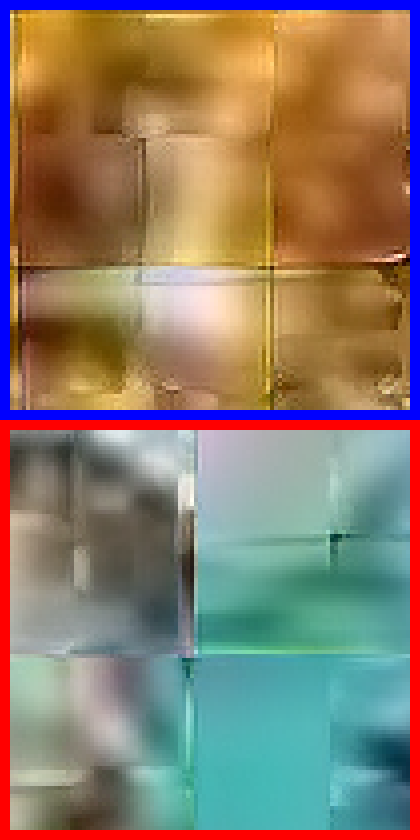}
        \includegraphics[width=\linewidth]{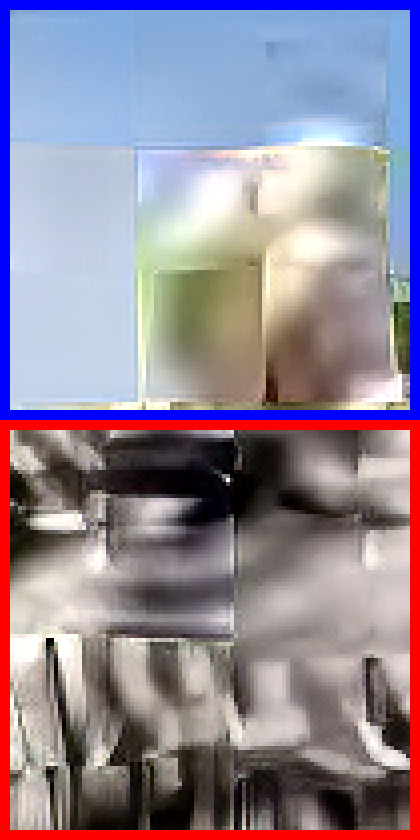}
        \centering \tiny ESRGAN \\ ~
        \vspace{0.2cm}
	\end{minipage}
	\hspace{-0.02\linewidth}
	\begin{minipage}[c]{0.098\linewidth}
        \includegraphics[width=\linewidth]{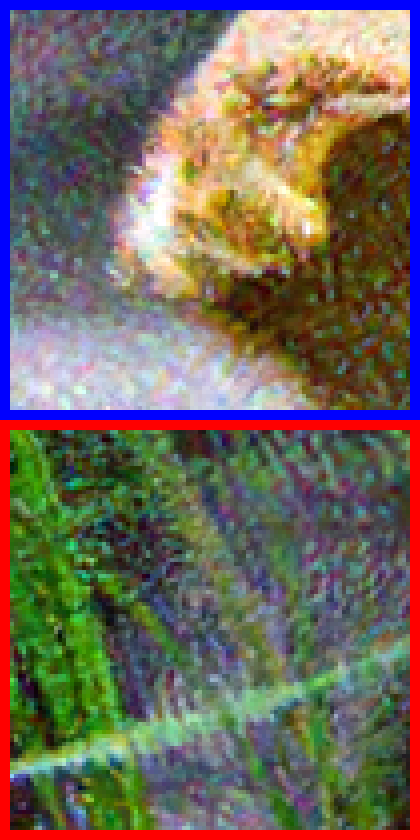}
        \includegraphics[width=\linewidth]{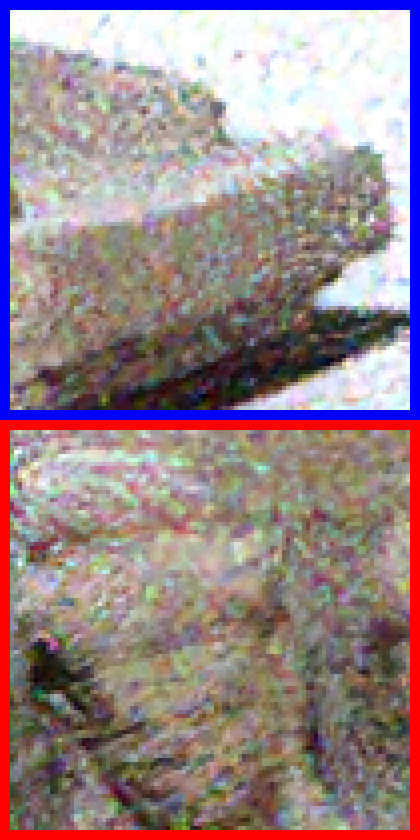}
        \includegraphics[width=\linewidth]{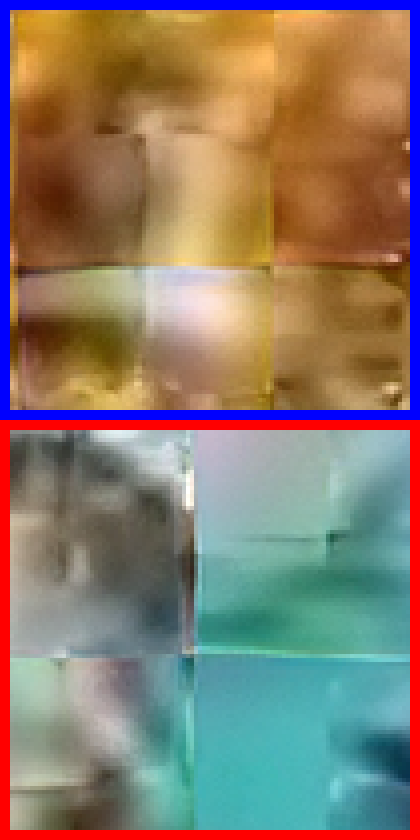}
        \includegraphics[width=\linewidth]{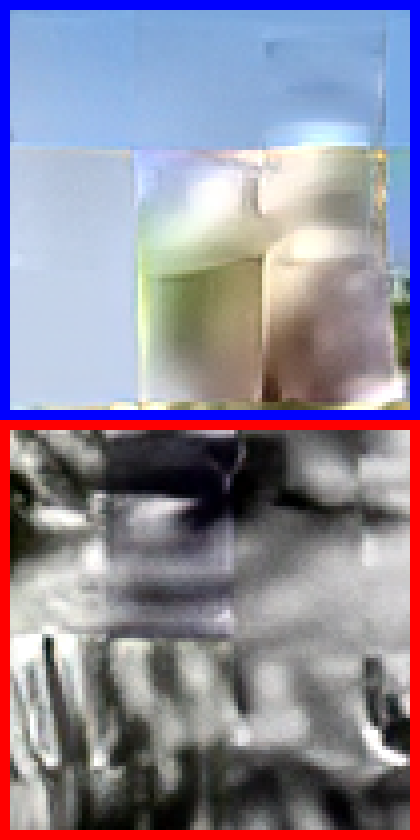}
        \centering \tiny Rank-\\SRGAN
        \vspace{0.2cm}
	\end{minipage}
	\hspace{-0.02\linewidth}
	\begin{minipage}[c]{0.098\linewidth}
        \includegraphics[width=\linewidth]{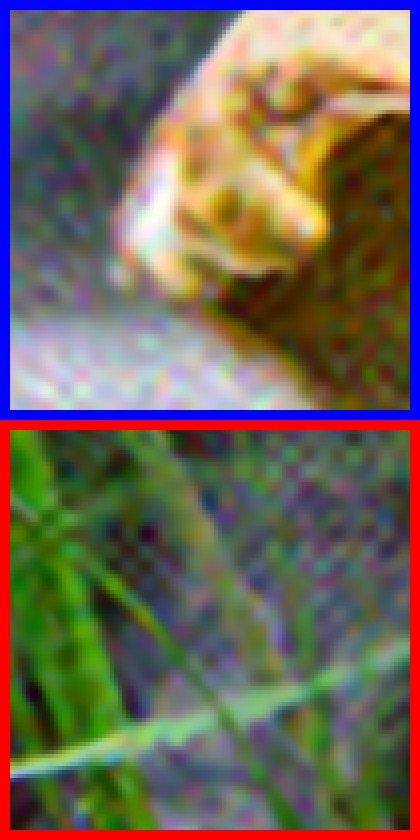}
        \includegraphics[width=\linewidth]{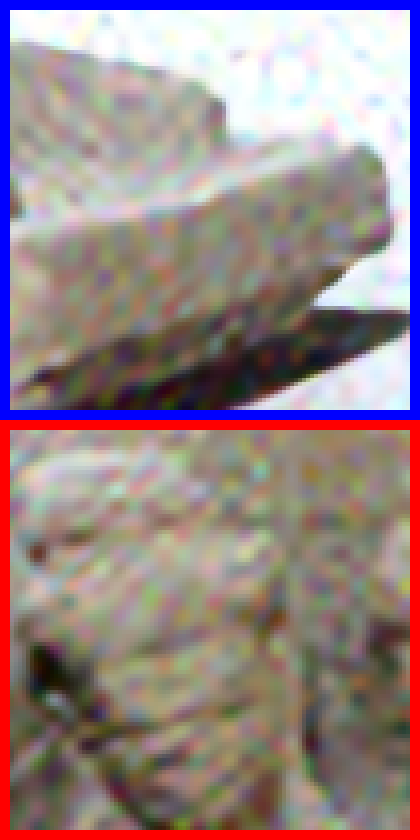}
        \includegraphics[width=\linewidth]{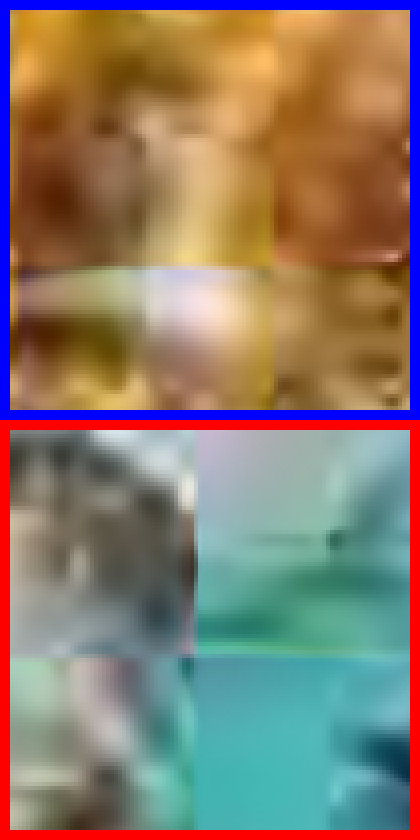}
        \includegraphics[width=\linewidth]{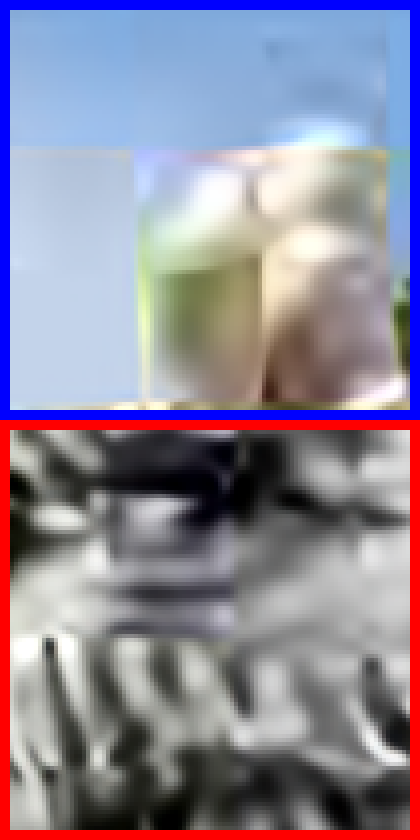}
        \centering \tiny EDSR \\ ~
        \vspace{0.2cm}
	\end{minipage}
	\hspace{-0.02\linewidth}
	\begin{minipage}[c]{0.098\linewidth}
        \includegraphics[width=\linewidth]{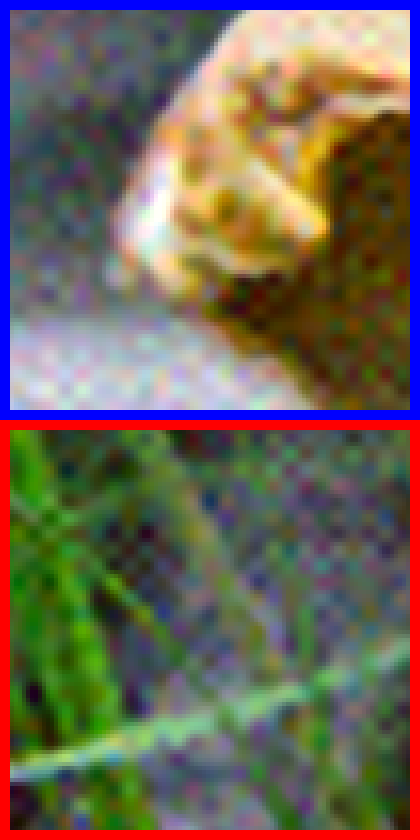}
        \includegraphics[width=\linewidth]{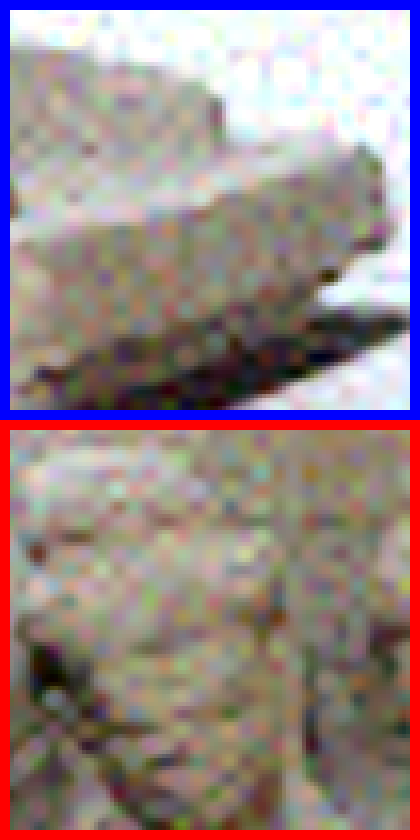}
        \includegraphics[width=\linewidth]{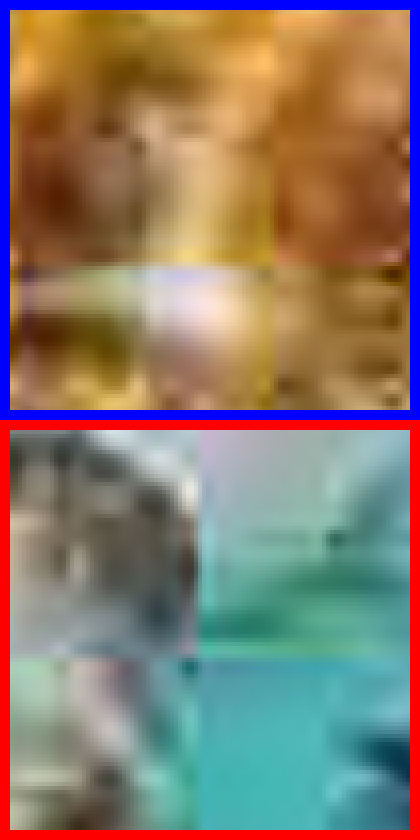}
        \includegraphics[width=\linewidth]{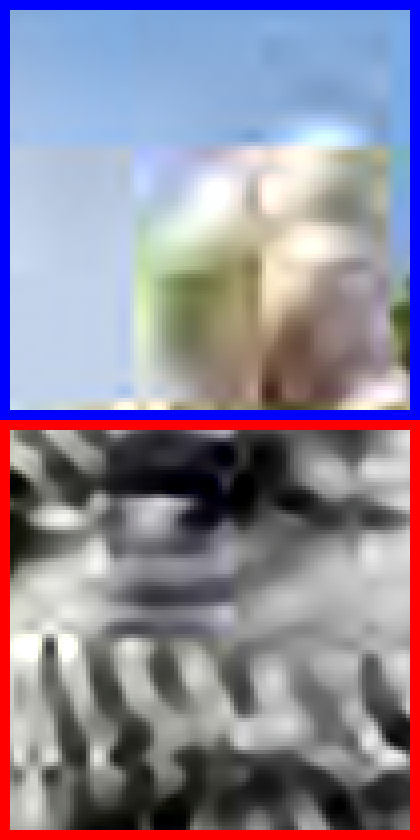}
        \centering \tiny ZSSR \\ ~
        \vspace{0.2cm}
	\end{minipage}
	\caption{Qualitative comparison of our approaches (bold) with other state-of-the-art methods on images with additional sensor noise (upper two rows) or corruption artifacts (lower two rows). \textbf{ours} refers to our ESRGAN-FS trained with data generated by our DSGAN method in the two different settings.}  
    \label{fig:corrupted_results}    
\end{figure}

\paragraph{Experiments on Corrupted Images}
\label{sec:experiments_corrupted}
In this experiment, our methods are fine-tuned on the corrupted DF2K datasets as described in Section~\ref{sec:experiments_datasets}. In Table~\ref{tab:experiments_corrupted}, we compare the PSNR, SSIM and LPIPS values of all methods on the DIV2K validation set. 
As ground truth, we use the corrupted and original HR images in the SDSR and TDSR settings, respectively. In our discussion of the quantitative evaluation, we focus on the LPIPS measure, as it has the best correlation with image similarity. A qualitative comparison is shown in Figure~\ref{fig:corrupted_results}. 
\begin{table}[!t]
	\centering
    \begin{sideways}
        \scriptsize \hspace{-0.16\linewidth}TDSR \hspace{0.11\linewidth} SDSR
    \end{sideways}
	\tabcolsep=0.01\linewidth
	\scriptsize
	\begin{tabular}{l|ccc|ccc}
		\multicolumn{1}{c}{ } & \multicolumn{3}{c}{sensor noise} & \multicolumn{3}{c}{compression artifacts} \\
	    method & PSNR$\uparrow$ & SSIM$\uparrow$ & LPIPS$\downarrow$ & PSNR$\uparrow$ & SSIM$\uparrow$ & LPIPS$\downarrow$ \\
		\hline
		\textbf{ours} & 20.87 & 0.39 & 0.26 & 22.03 & 0.57 & 0.35 \\
		ESRGAN (FT) & 18.32 & 0.22 & 0.64 & 23.29 & 0.62 & 0.42 \\
		ESRGAN~\cite{WangYWGLDQL18} & 17.18 & 0.19 & 0.74 & 22.60 & 0.60 & 0.47 \\
		RankSRGAN~\cite{zhang2019ranksrgan} & 19.78 & 0.27 & 0.56 & 23.34 & 0.61 & 0.43 \\
		EDSR~\cite{LimSKNL17} & 23.40 & 0.44 & 0.71 & 23.96 & 0.64 & 0.45 \\
		ZSSR~\cite{ShocherCI18} & 23.18 & 0.43 & 0.73 & 24.02 & 0.64 & 0.46 \\
		\hline
		\textbf{ours} & 22.52 & 0.52 & 0.33 & 20.39 & 0.50 & 0.42 \\
		ESRGAN (FT) & 18.61 & 0.24 & 0.81 & 23.10 & 0.60 & 0.50 \\
		ESRGAN~\cite{WangYWGLDQL18} & 17.39 & 0.19 & 0.94 & 22.43 & 0.58 & 0.53 \\
		RankSRGAN~\cite{zhang2019ranksrgan} & 20.19 & 0.30 & 0.68 & 23.16 & 0.59 & 0.49 \\
		EDSR~\cite{LimSKNL17} & 24.48 & 0.53 & 0.68 & 23.75 & 0.62 & 0.54 \\
		ZSSR~\cite{ShocherCI18} & 24.21 & 0.52 & 0.70 & 23.81 & 0.62 & 0.55 \\
	\end{tabular}
    \caption{We report the mean scores over the DIV2K validation set with added corruptions. The arrows indicate if high $\uparrow$ or low $\downarrow$ values are desired. \textbf{ours} refers to our ESRGAN-FS trained with data generated by our DSGAN method.}
	\label{tab:experiments_corrupted}
\end{table}

All state-of-the-art methods introduce severe artifacts in the output for both sensor noise and compression artifacts. Specifically, ESRGAN achieves the worst LPIPS value in both the TDSR and SDSR settings. This value can be improved through fine-tuning on the corrupted images. However, the visual quality does not show any notable improvements. RankSRGAN achieves the best LPIPS value of the state-of-the-art methods, but still introduces significant corruptions in the output. EDSR and ZSSR manage to reduce the corruptions as they generally produce more blurry results. However, the LPIPS value is still close to the value for ESRGAN. In contrast, our methods produce satisfying results in all cases. This is reflected in the LPIPS values, as well as the visual quality of the images, which show almost no corruptions. This is due to our model being trained with input images with similar characteristics as the validation images that were upsampled to generate these results.

\begin{figure}[!t]
    \centering  
	\begin{minipage}[c]{0.22\linewidth}
	    \includegraphics[width=\linewidth]{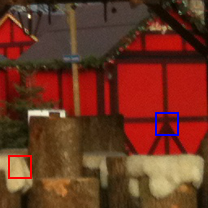}
	    \includegraphics[width=\linewidth]{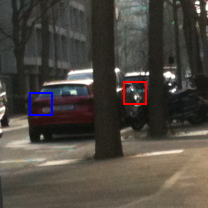}
	    \includegraphics[width=\linewidth]{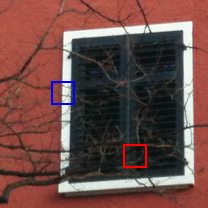}
	    \includegraphics[width=\linewidth]{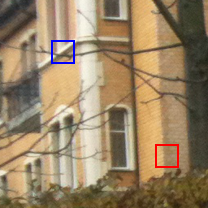}
	    \centering \tiny original \\ ~
	    \vspace{0.2cm}
	\end{minipage}
	\hspace{-0.02\linewidth}
	\begin{minipage}[c]{0.11\linewidth}
        \includegraphics[width=\linewidth]{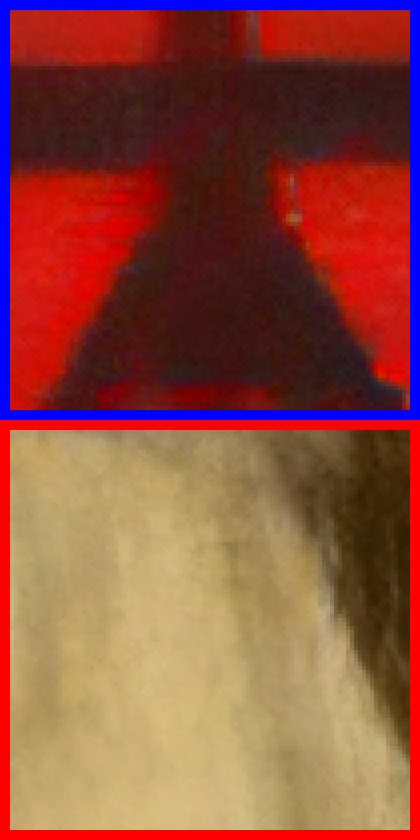}
        \includegraphics[width=\linewidth]{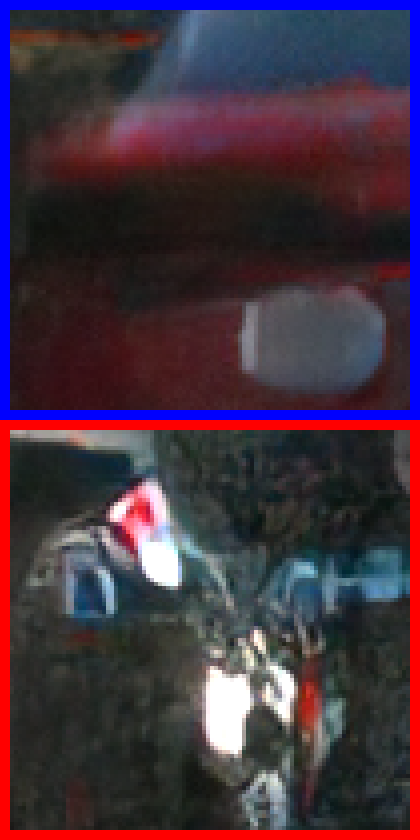}
        \includegraphics[width=\linewidth]{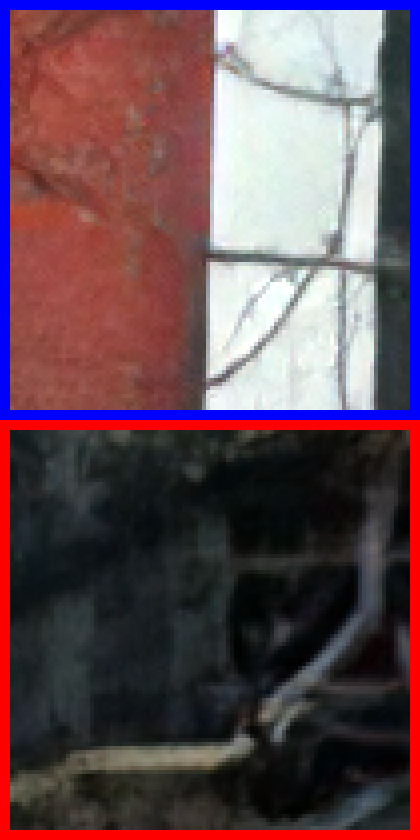}
        \includegraphics[width=\linewidth]{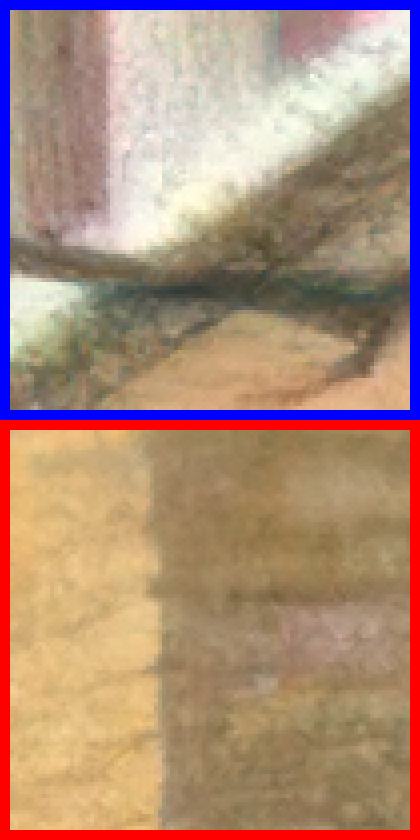}
        \vspace{-0.03cm}
        \centering \tiny \textbf{ours} \\ (TDSR)
        \vspace{0.2cm}
	\end{minipage}
	\hspace{-0.02\linewidth}
	\begin{minipage}[c]{0.11\linewidth}
        \includegraphics[width=\linewidth]{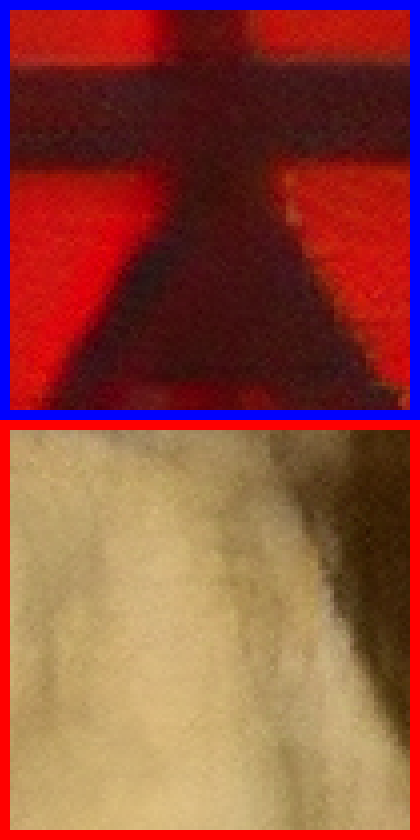}
        \includegraphics[width=\linewidth]{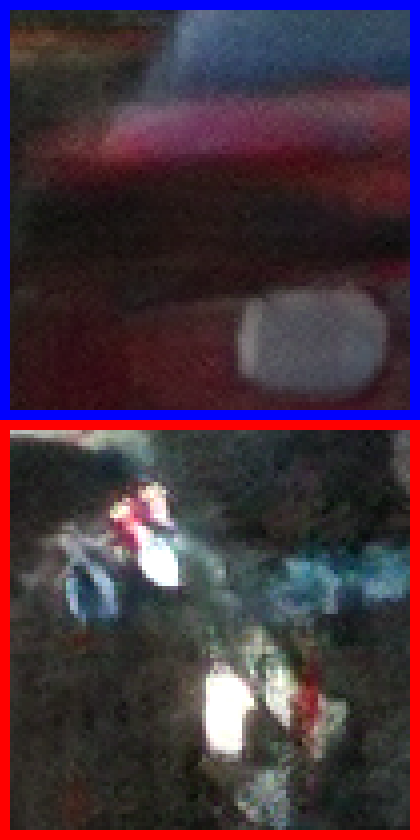}
        \includegraphics[width=\linewidth]{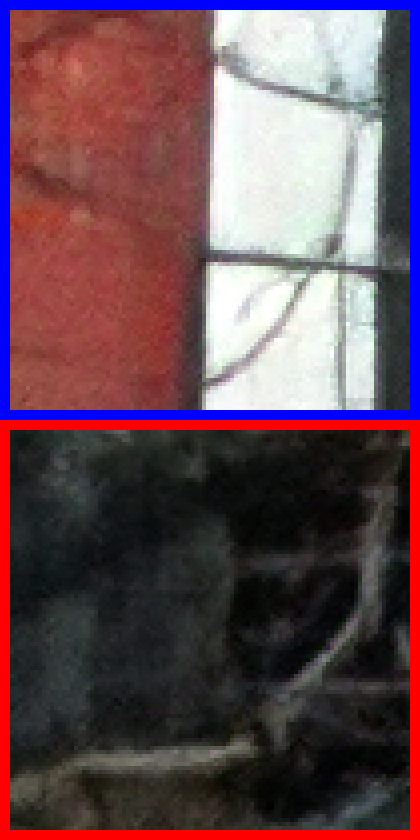}
        \includegraphics[width=\linewidth]{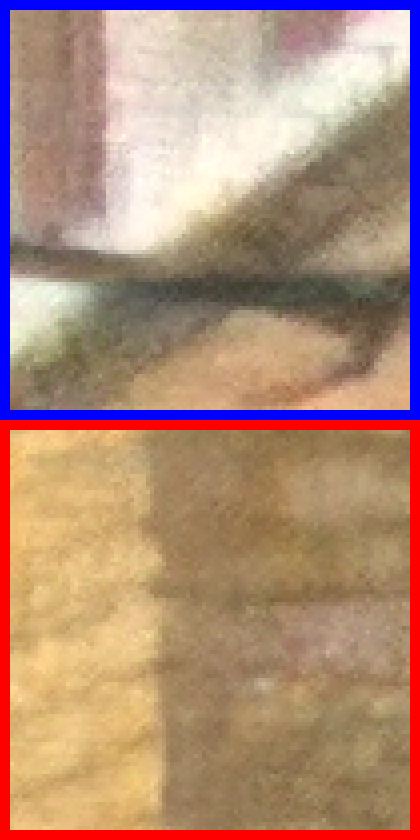}
        \vspace{-0.03cm}
        \centering \tiny \textbf{ours} \\ (SDSR)
        \vspace{0.2cm}
	\end{minipage}
	\hspace{-0.02\linewidth}
	\begin{minipage}[c]{0.11\linewidth}
        \includegraphics[width=\linewidth]{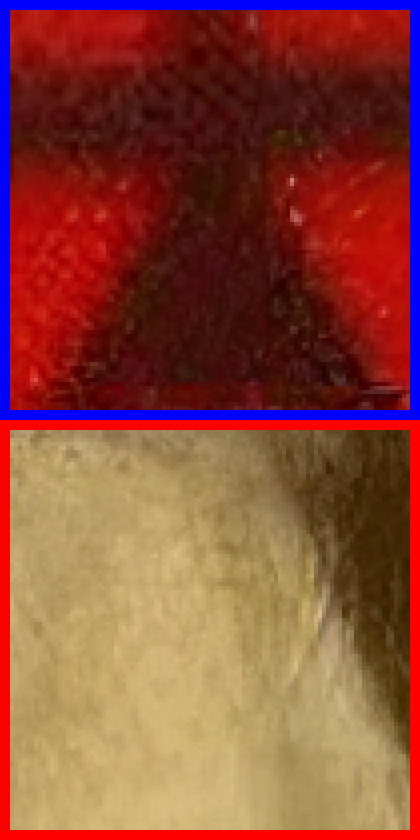}
        \includegraphics[width=\linewidth]{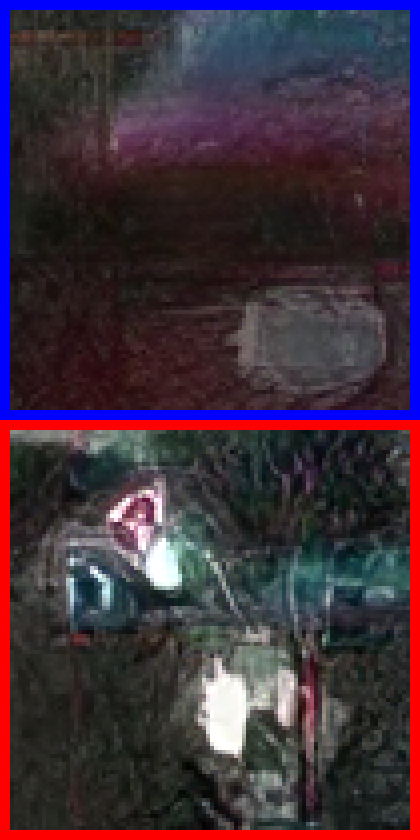}
        \includegraphics[width=\linewidth]{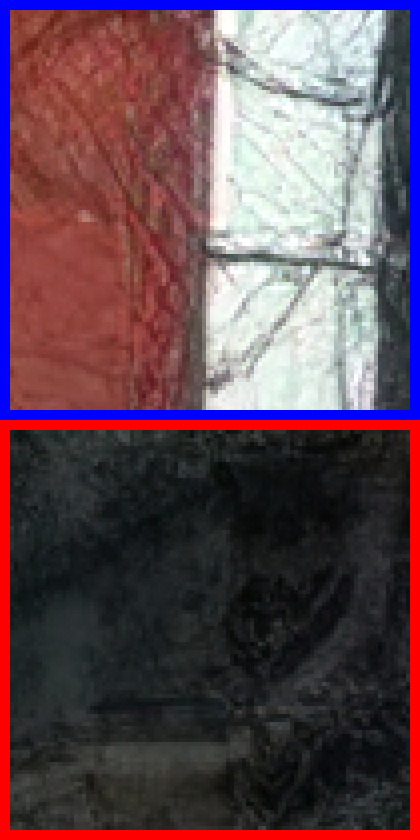}
        \includegraphics[width=\linewidth]{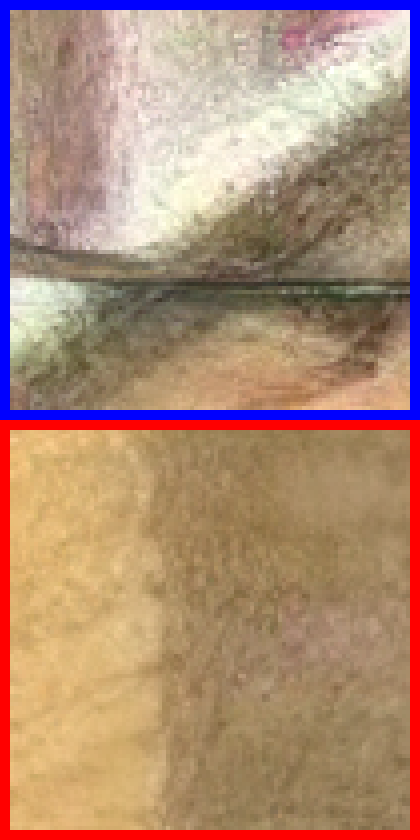}
        \vspace{-0.03cm}    
        \centering \tiny ESRGAN \\ (FT)
        \vspace{0.2cm}
	\end{minipage}
	\hspace{-0.02\linewidth}
	\begin{minipage}[c]{0.11\linewidth}
        \includegraphics[width=\linewidth]{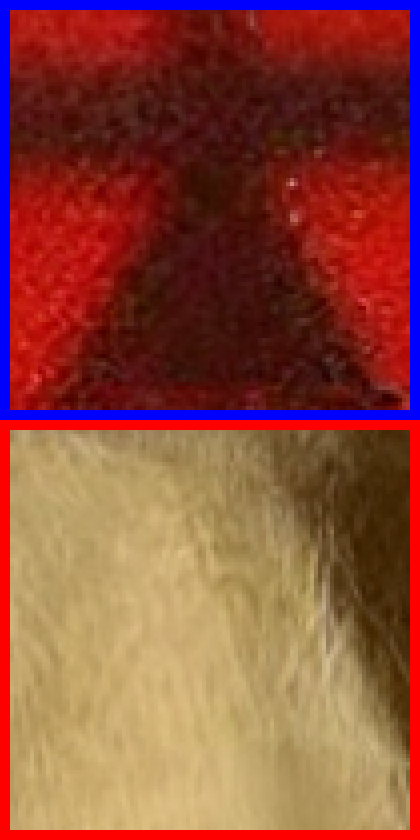}
        \includegraphics[width=\linewidth]{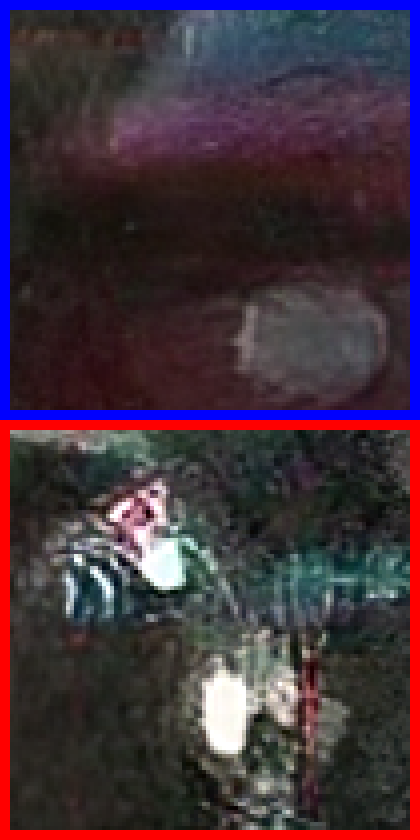}
        \includegraphics[width=\linewidth]{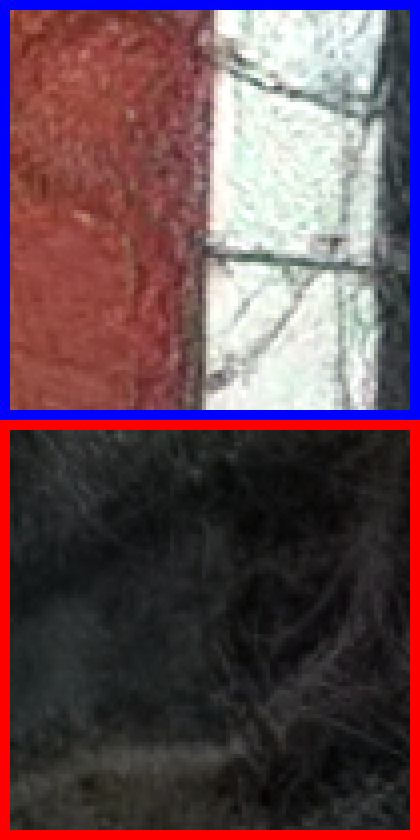}
        \includegraphics[width=\linewidth]{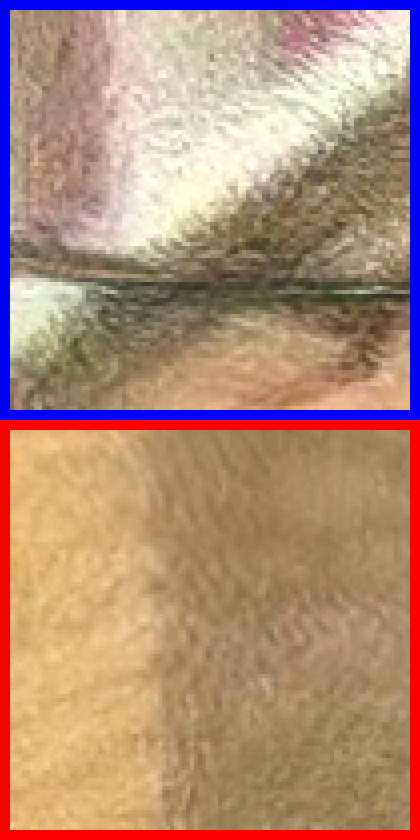}
        \centering \tiny ESRGAN \\ ~
        \vspace{0.2cm}
	\end{minipage}
	\hspace{-0.02\linewidth}
	\begin{minipage}[c]{0.11\linewidth}
        \includegraphics[width=\linewidth]{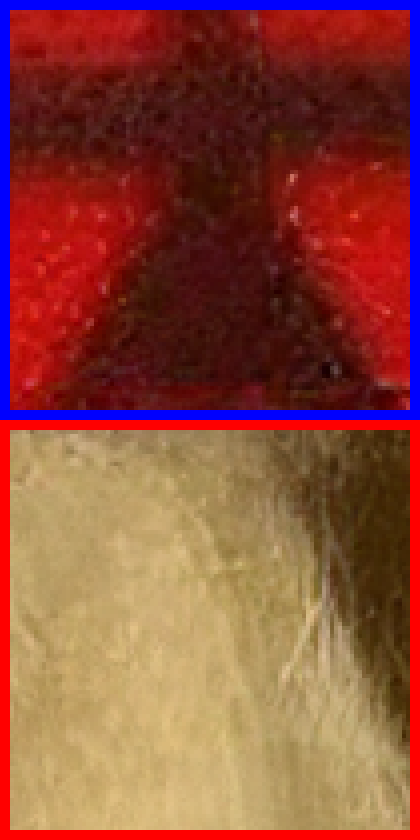}
        \includegraphics[width=\linewidth]{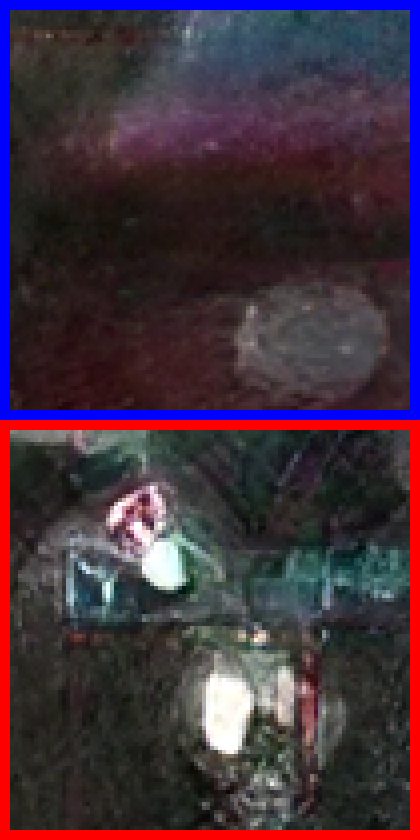}
        \includegraphics[width=\linewidth]{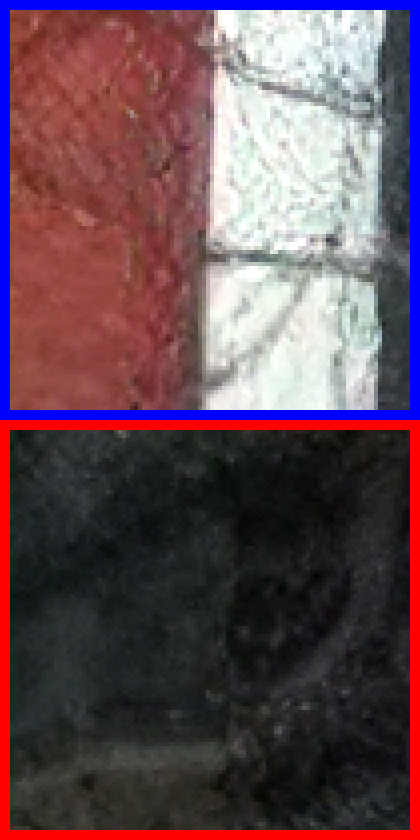}
        \includegraphics[width=\linewidth]{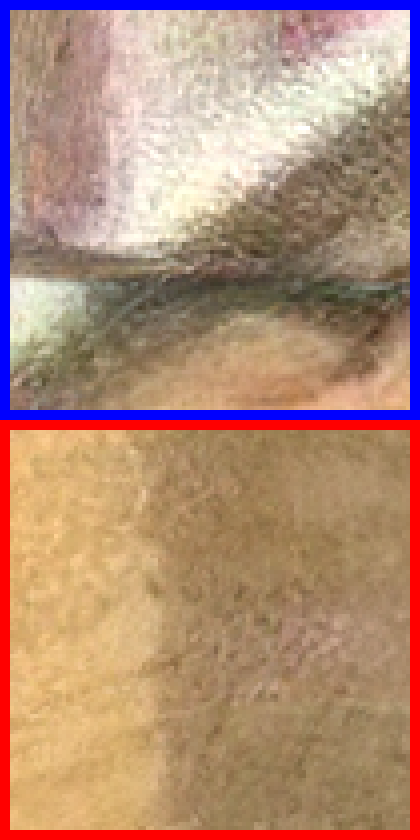}
        \centering \tiny Rank-\\SRGAN
        \vspace{0.2cm}
	\end{minipage}
	\hspace{-0.02\linewidth}
	\begin{minipage}[c]{0.11\linewidth}
        \includegraphics[width=\linewidth]{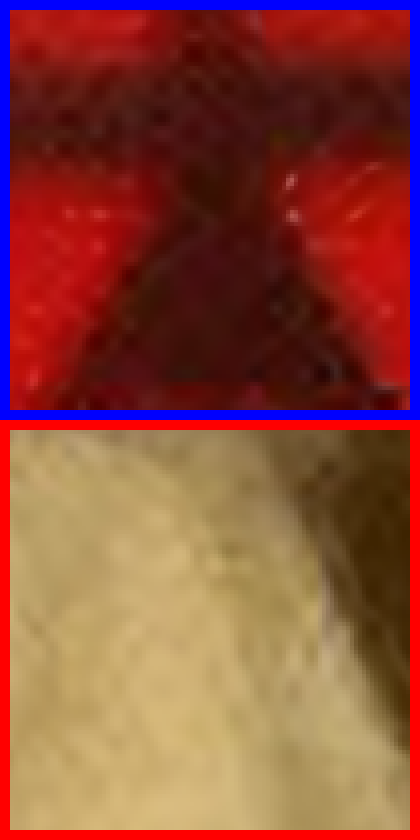}
        \includegraphics[width=\linewidth]{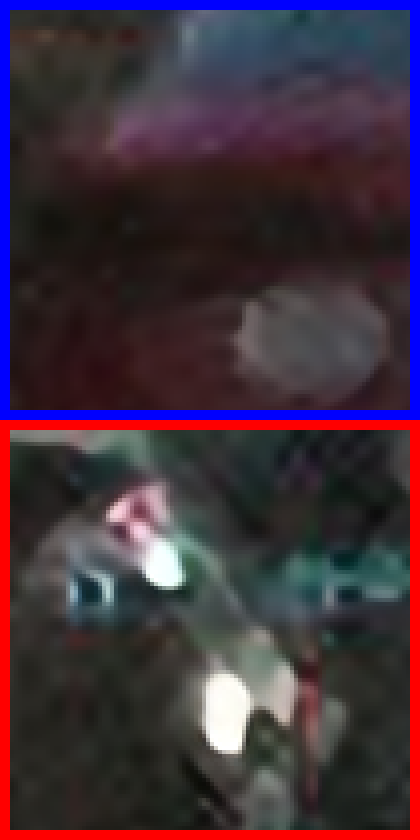}
        \includegraphics[width=\linewidth]{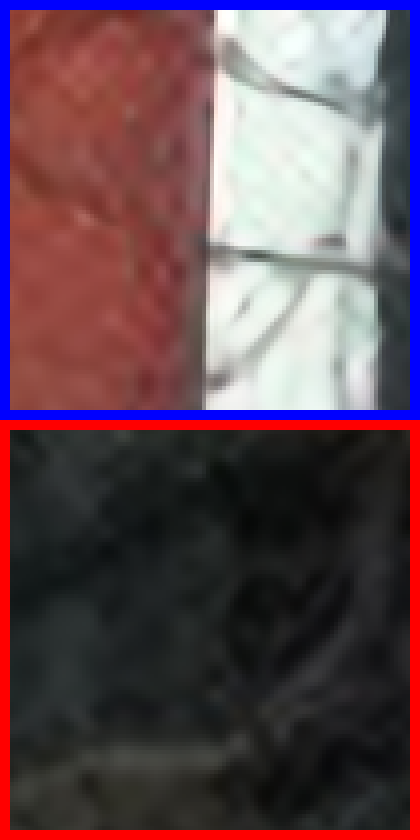}
        \includegraphics[width=\linewidth]{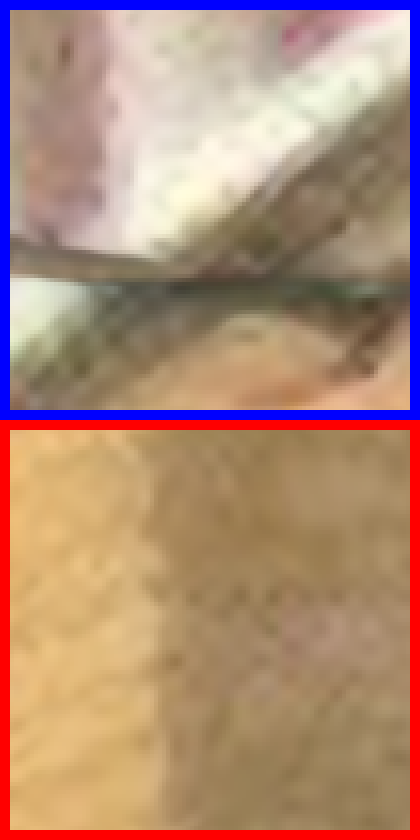}
        \centering \tiny EDSR \\ ~
        \vspace{0.2cm}
	\end{minipage}
	\hspace{-0.02\linewidth}
	\begin{minipage}[c]{0.11\linewidth}
        \includegraphics[width=\linewidth]{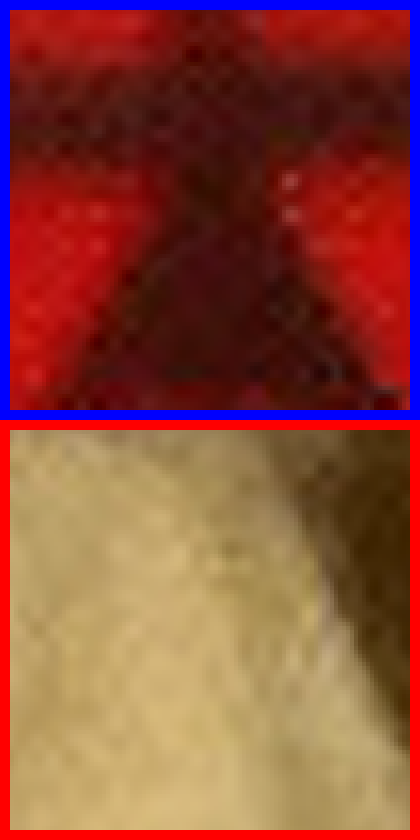}
        \includegraphics[width=\linewidth]{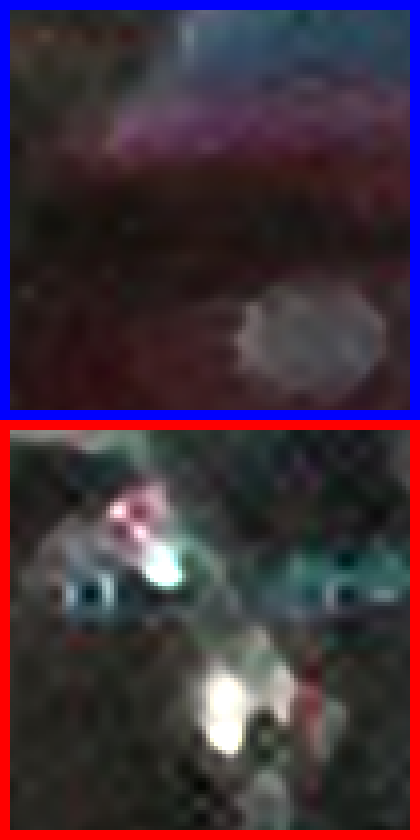}
        \includegraphics[width=\linewidth]{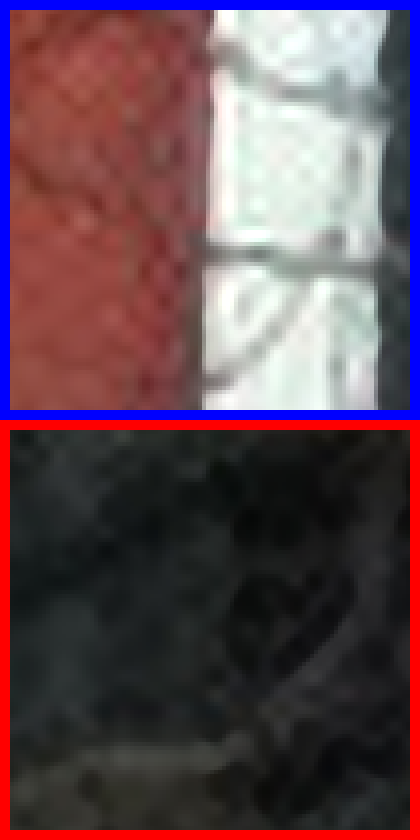}
        \includegraphics[width=\linewidth]{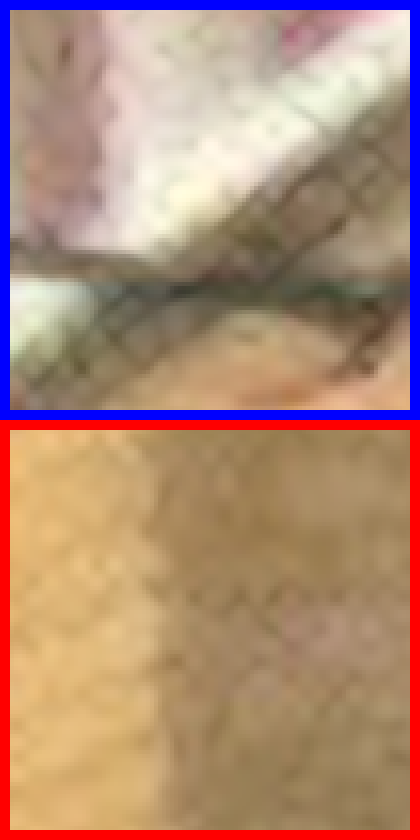}
        \centering \tiny ZSSR \\ ~
        \vspace{0.2cm}
	\end{minipage}
    \caption{Qualitative comparison of our approaches (bold) with other state-of-the-art methods on real-world images from the DPED dataset taken with an iPhone 3 camera. \textbf{ours} refers to our ESRGAN-FS trained with data generated by our DSGAN method in the two different settings.}
    \label{fig:real_world_results}    
\end{figure}

\paragraph{Experiments on Real-World Images}
\label{sec:experiments_real_world}
We also evaluate our method on real-world images from the DPED dataset~\cite{ignatov2017dslr}, where we use the train images captured by an iPhone 3 camera for fine-tuning. 
Since we do not have access to ground truth, we only compare the results visually, which is done in Figure~\ref{fig:real_world_results}. Due to the significant amount of corruptions (cf. Figure~\ref{fig:real_world_bicubic}), none of the state-of-the-art methods produces satisfying results on this dataset. ESRGAN and RankSRGAN introduce strong artifacts, which are slightly reduced in the PSNR based EDSR. ZSSR produces visually very similar results, which are rather blurry. In contrast, our models produce images that are sharp and greatly reduce the amount of corruptions.

\paragraph{The AIM 2019 Challenge}
\label{sec:experiments_aim}
We also participated in the AIM 2019 Real World Super-Resolution Challenge~\cite{AIM2019RWSRchallenge}. Thereby, the SR ($\times4$) images are supposed to either match the source domain in \textit{Same Domain Real World Super-Resolution (SDSR)} or a clean target domain in \textit{Target Domain Real World Super-Resolution (TDSR)}. 

We first train our DSGAN model on the corrupted source dataset. To increase the diversity in the data, we randomly flip and rotate the images. 
For \textit{SDSR}, our HR dataset contains the source and target datasets. Since the latter contains clean images, we use DSGAN to add the corruptions. 
In \textit{TDSR}, the HR dataset is constructed by combining the target dataset and the bicubically downscaled source dataset, for which we use a scaling factor of $2$. 
In both cases, the LR images are created by applying DSGAN on the HR images.
We then fine-tune ESRGAN-FS on these datasets.

\begin{figure}[!t]
    \centering  
	\begin{minipage}[c]{0.22\linewidth}
	    \includegraphics[width=\linewidth]{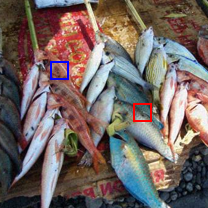}
	    \includegraphics[width=\linewidth]{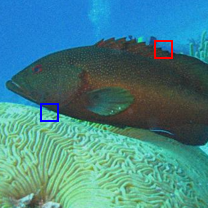}
	    \includegraphics[width=\linewidth]{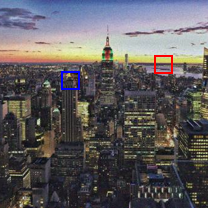}
	    \includegraphics[width=\linewidth]{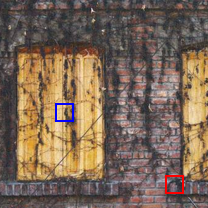}
	    \centering \tiny original \\ ~
	    \vspace{0.2cm}
	\end{minipage}
	\hspace{-0.02\linewidth}
	\begin{minipage}[c]{0.11\linewidth}
        \includegraphics[width=\linewidth]{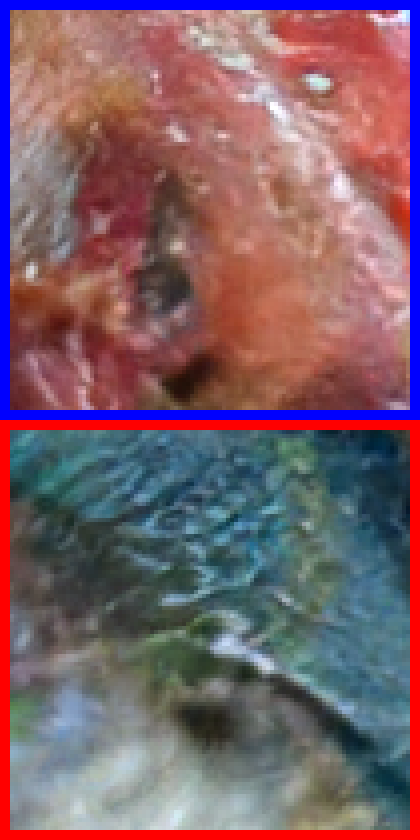}
        \includegraphics[width=\linewidth]{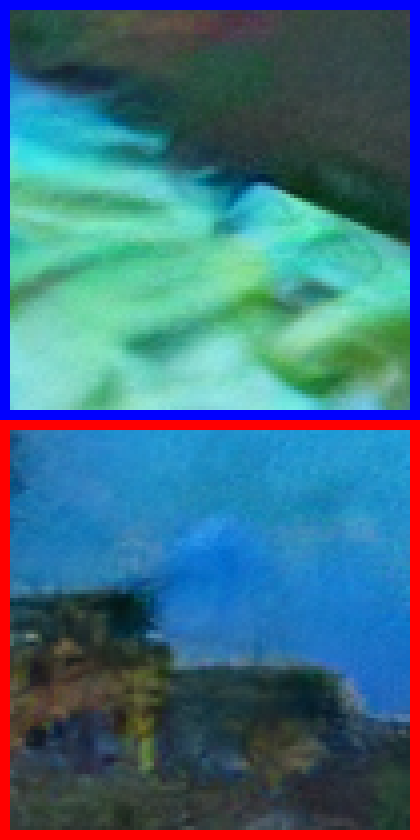}
        \includegraphics[width=\linewidth]{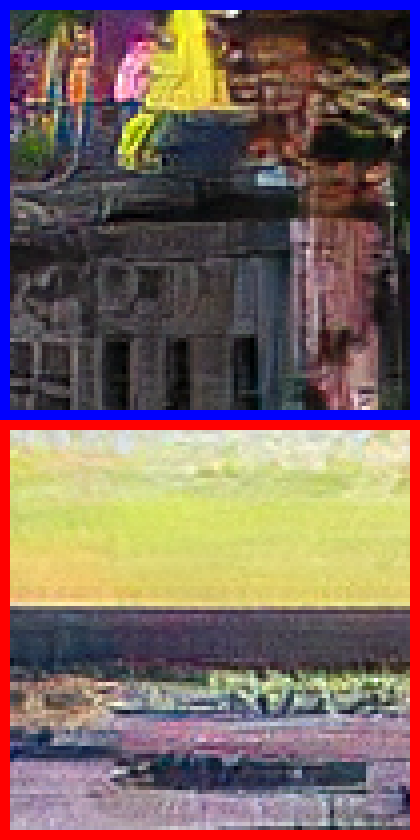}
        \includegraphics[width=\linewidth]{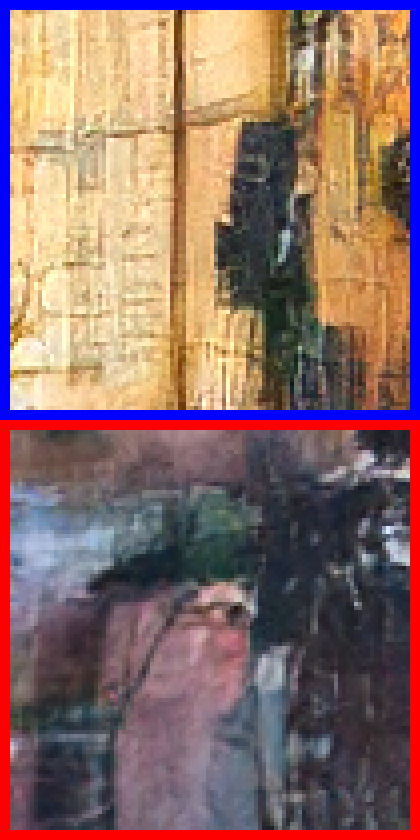}
        \vspace{-0.03cm}
        \centering \tiny \textbf{ours} \\ (TDSR)
        \vspace{0.2cm}
	\end{minipage}
	\hspace{-0.02\linewidth}
	\begin{minipage}[c]{0.11\linewidth}
        \includegraphics[width=\linewidth]{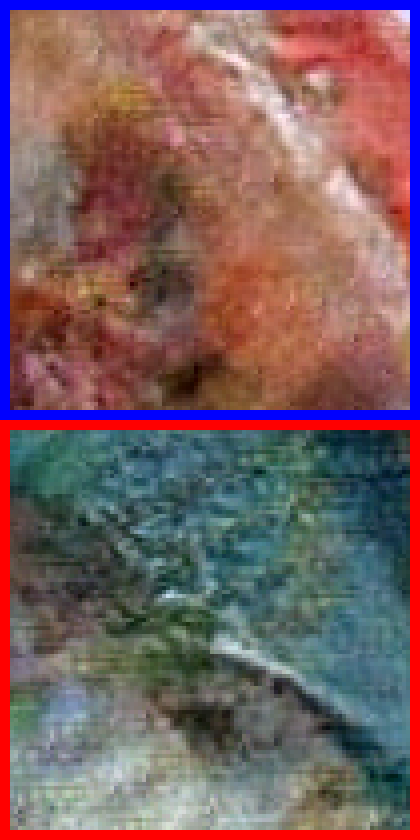}
        \includegraphics[width=\linewidth]{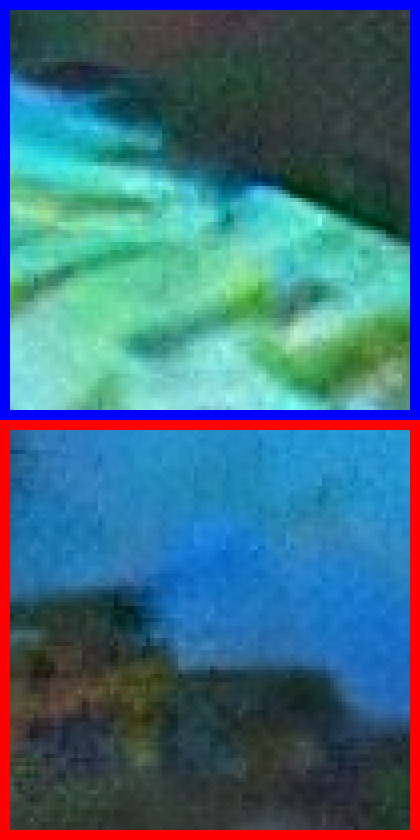}
        \includegraphics[width=\linewidth]{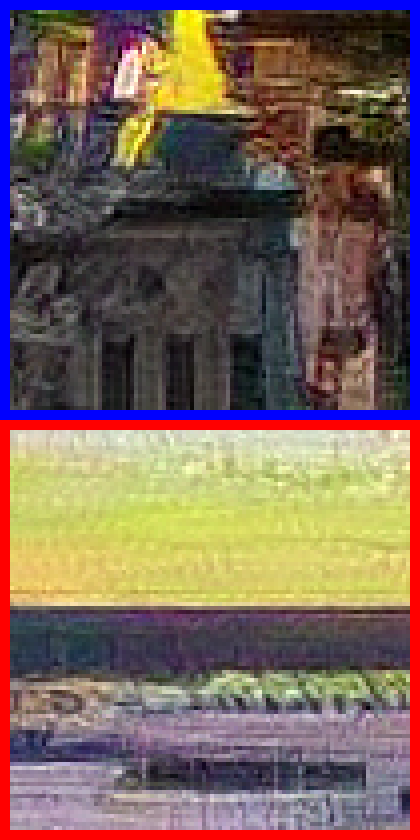}
        \includegraphics[width=\linewidth]{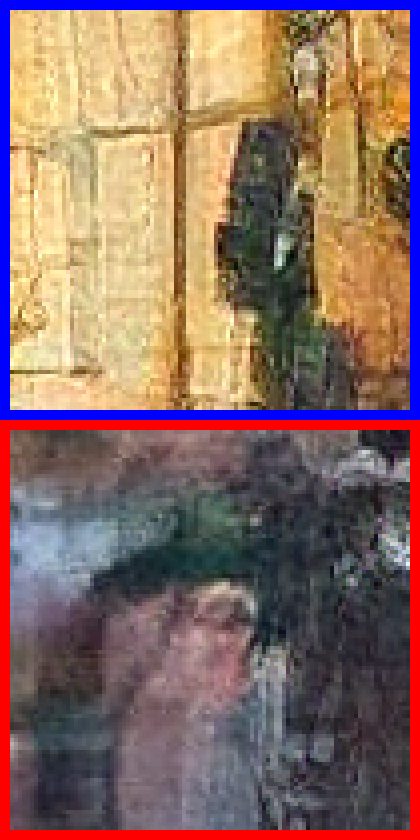}
        \vspace{-0.03cm}
        \centering \tiny \textbf{ours} \\ (SDSR)
        \vspace{0.2cm}
	\end{minipage}
	\hspace{-0.02\linewidth}
	\begin{minipage}[c]{0.11\linewidth}
        \includegraphics[width=\linewidth]{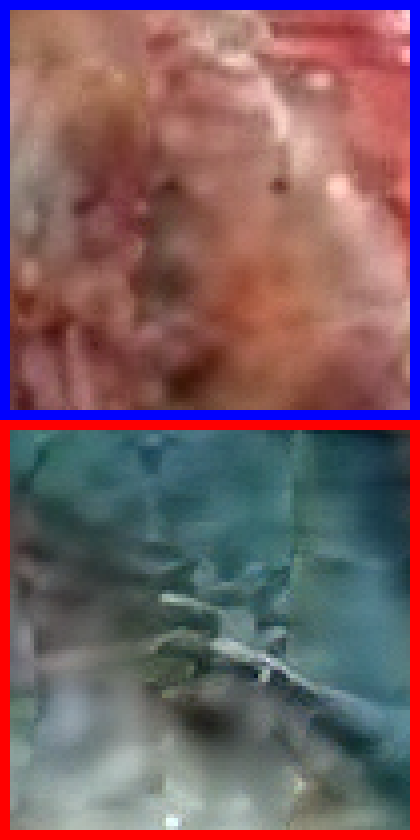}
        \includegraphics[width=\linewidth]{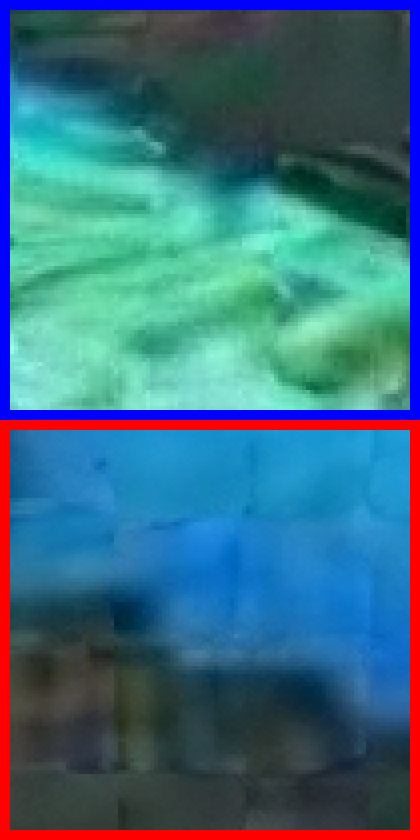}
        \includegraphics[width=\linewidth]{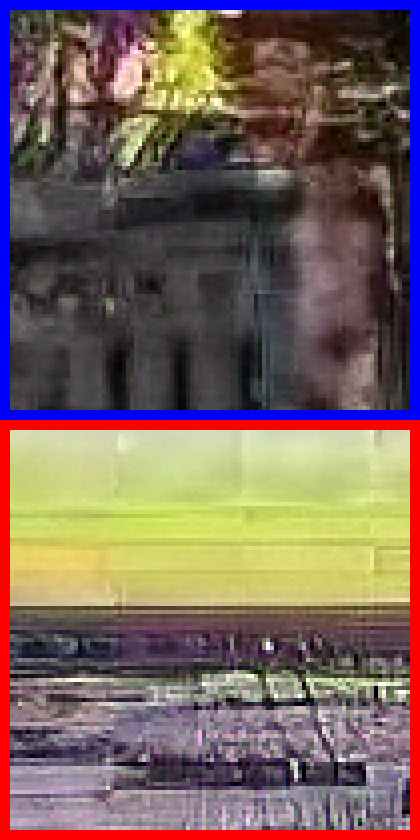}
        \includegraphics[width=\linewidth]{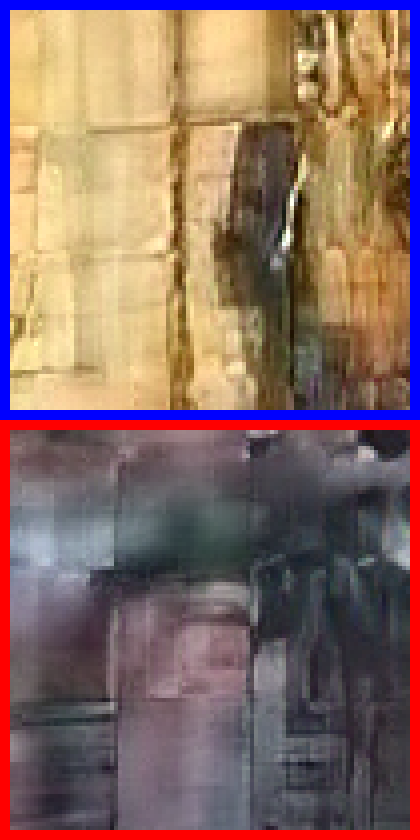}
        \vspace{-0.03cm}    
        \centering \tiny ESRGAN \\ (FT)
        \vspace{0.2cm}
	\end{minipage}
	\hspace{-0.02\linewidth}
	\begin{minipage}[c]{0.11\linewidth}
        \includegraphics[width=\linewidth]{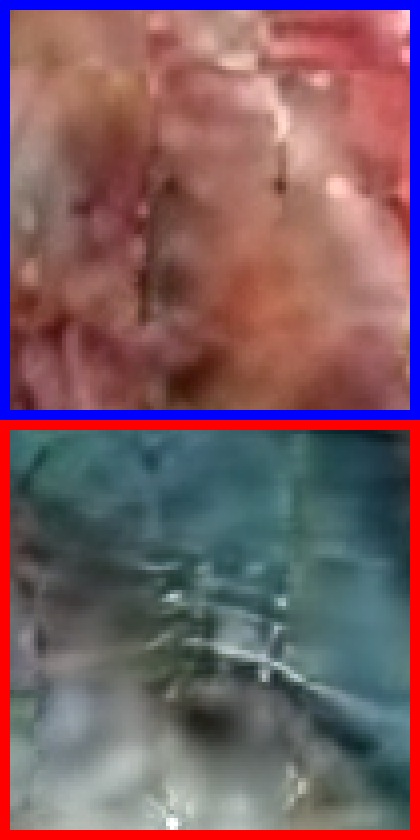}
        \includegraphics[width=\linewidth]{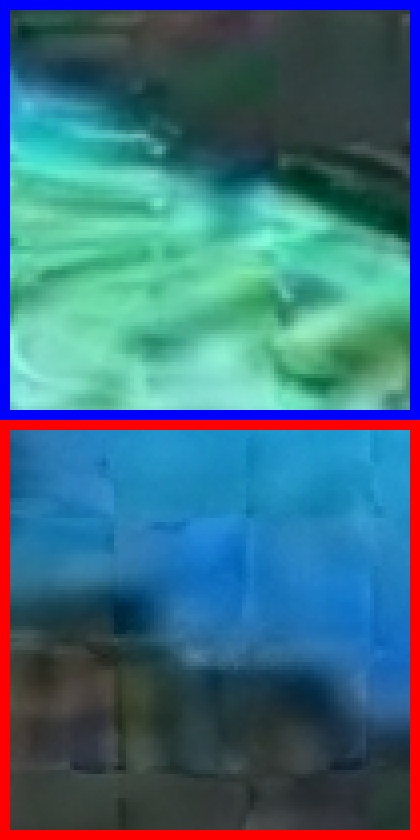}
        \includegraphics[width=\linewidth]{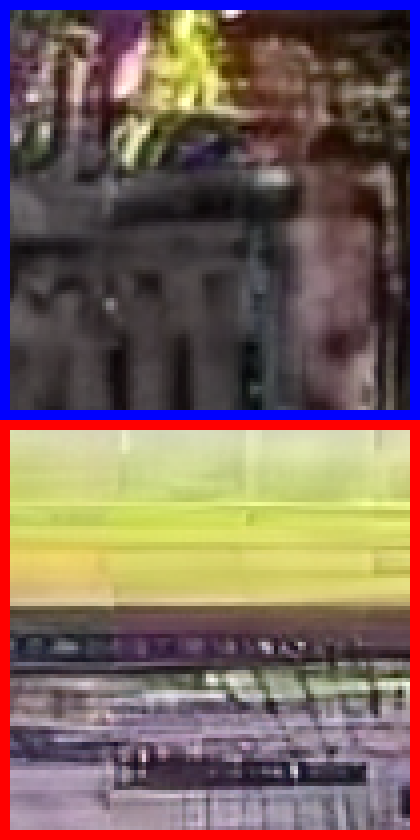}
        \includegraphics[width=\linewidth]{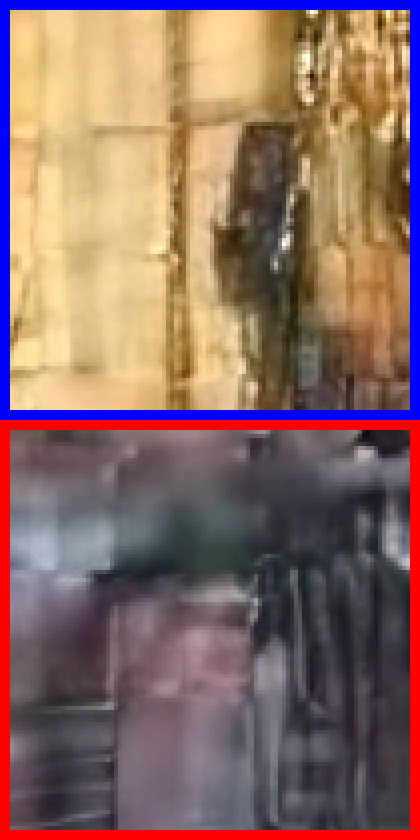}
        \centering \tiny ESRGAN \\ ~
        \vspace{0.2cm}
	\end{minipage}
	\hspace{-0.02\linewidth}
	\begin{minipage}[c]{0.11\linewidth}
        \includegraphics[width=\linewidth]{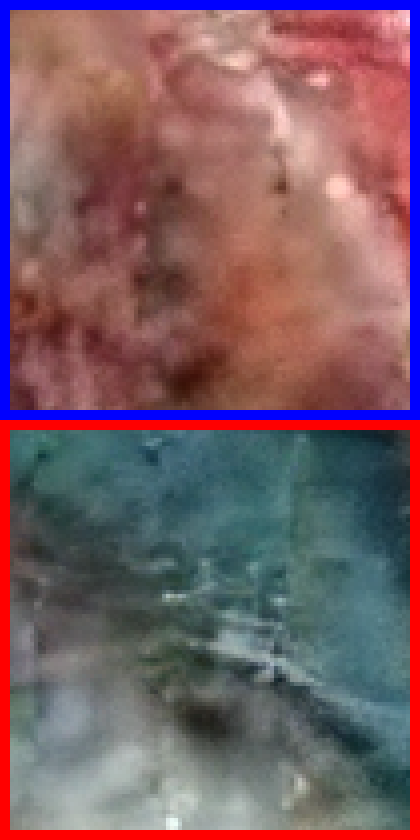}
        \includegraphics[width=\linewidth]{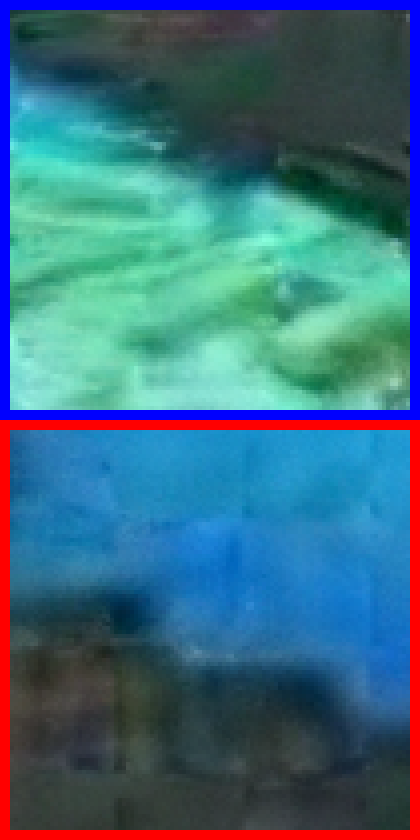}
        \includegraphics[width=\linewidth]{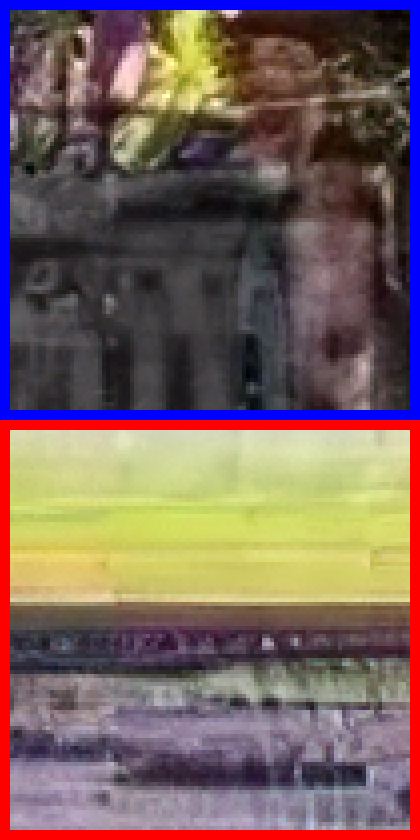}
        \includegraphics[width=\linewidth]{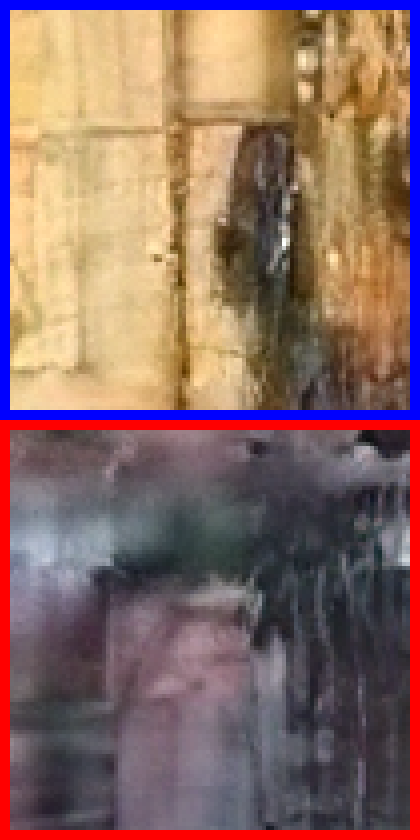}
        \centering \tiny Rank-\\SRGAN
        \vspace{0.2cm}
	\end{minipage}
	\hspace{-0.02\linewidth}
	\begin{minipage}[c]{0.11\linewidth}
        \includegraphics[width=\linewidth]{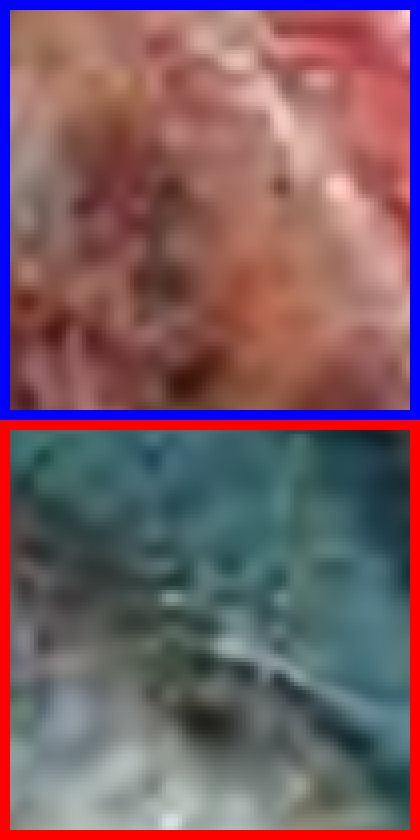}
        \includegraphics[width=\linewidth]{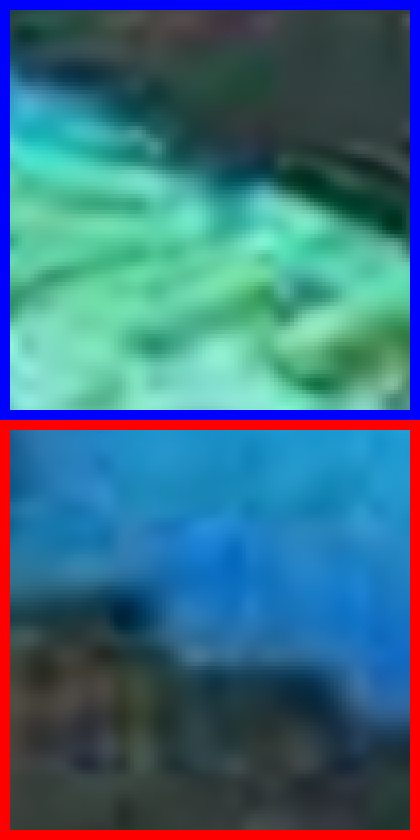}
        \includegraphics[width=\linewidth]{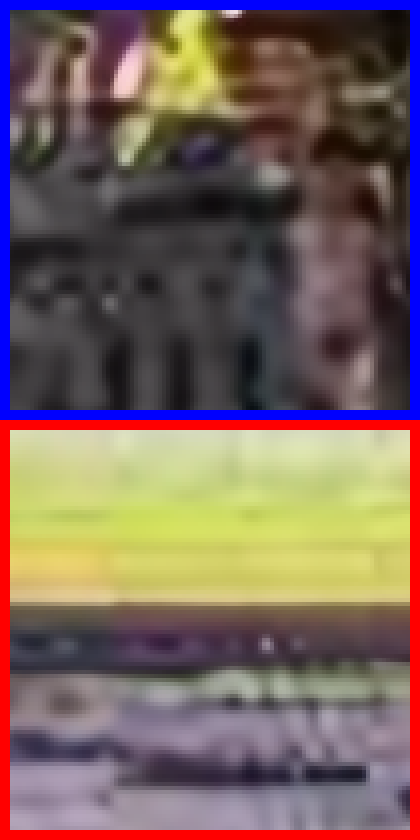}
        \includegraphics[width=\linewidth]{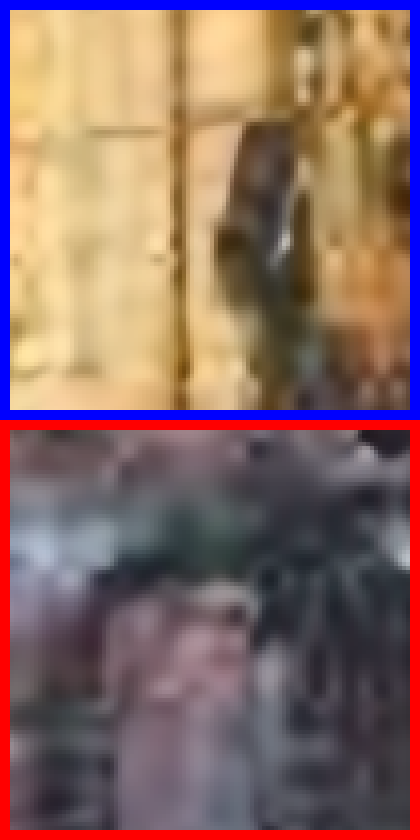}
        \centering \tiny EDSR \\ ~
        \vspace{0.2cm}
	\end{minipage}
	\hspace{-0.02\linewidth}
	\begin{minipage}[c]{0.11\linewidth}
        \includegraphics[width=\linewidth]{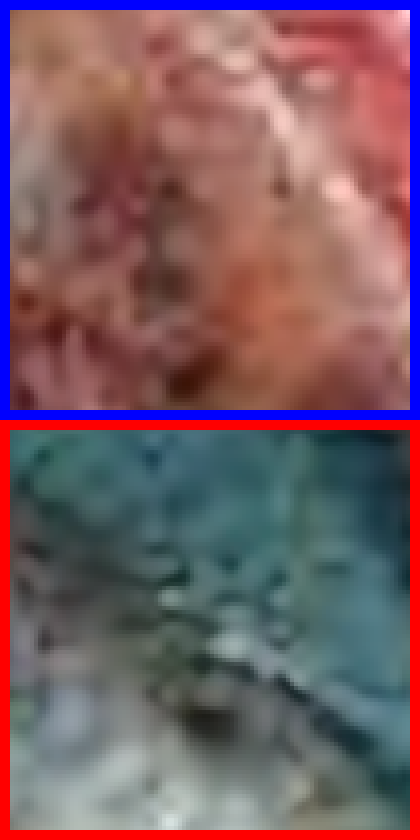}
        \includegraphics[width=\linewidth]{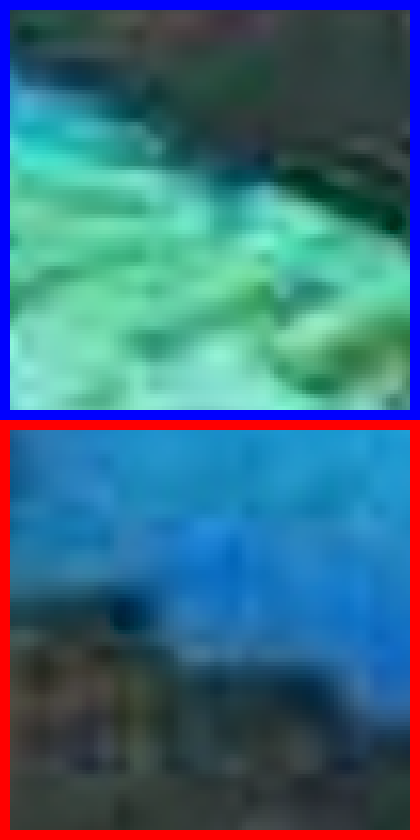}
        \includegraphics[width=\linewidth]{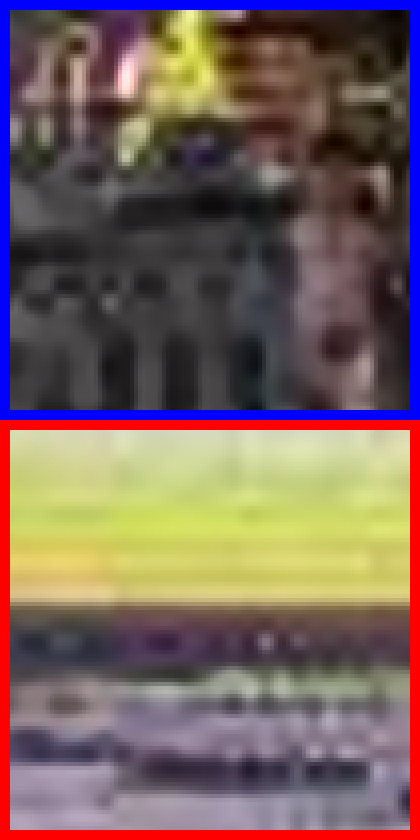}
        \includegraphics[width=\linewidth]{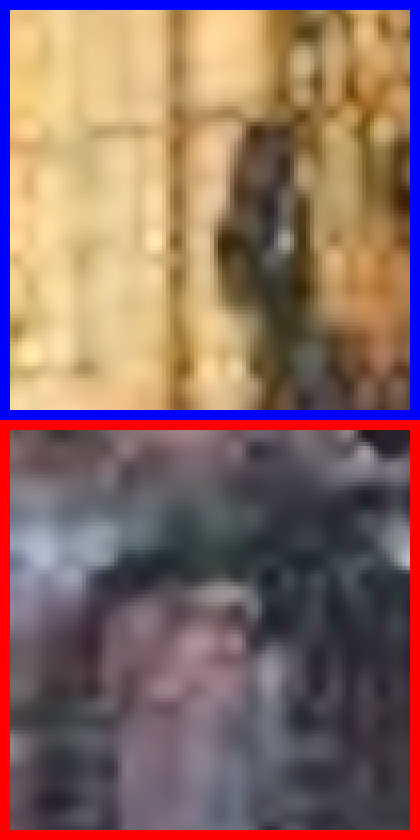}
        \centering \tiny ZSSR \\ ~
        \vspace{0.2cm}
	\end{minipage}
    \caption{Qualitative comparison of our approaches (bold) with other state-of-the-art methods on the AIM2019 Real World SR test set. \textbf{ours} refers to our ESRGAN-FS trained with data generated by our DSGAN method in the two different settings.}  
    \label{fig:aim_results}    
\end{figure}

In Figure~\ref{fig:aim_results}, we provide qualitative results for our SDSR and TDSR method and compare them to other state-of-the-art methods. As expected, PSNR based methods like EDSR~\cite{LimSKNL17} produce blurry results. Perception-driven methods, such as ESRGAN~\cite{WangYWGLDQL18}, generate sharper images, but they also increase the effect of image corruptions. On the other hand, our method produces sharp images with only few corruptions in both the SDSR and TDSR case. Most notably, the block structure corruptions caused by compression artifacts in the input image are removed in our models.

Furthermore, we present the results from the challenge in Table~\ref{tab:aim_results}. In addition to PSNR, SSIM and LPIPS, a Mean-Opinion-Score (MOS) was conducted to evaluate how similar the results are to ground truth. For both SDSR and TDSR our method won the challenge by achieving the lowest MOS. More information on the evaluation and competing methods can be found in the challenge report \cite{AIM2019RWSRchallenge}.

\begin{table}[!t]
	\centering
	\begin{sideways}
	\scriptsize \hspace{-0.17\linewidth}TDSR \hspace{0.1\linewidth} SDSR
	\end{sideways}
	\tabcolsep=0.01\linewidth
	\scriptsize
	\begin{tabular}{l|cccc}
		Method & PSNR$\uparrow$ & SSIM$\uparrow$ & LPIPS$\downarrow$ & MOS$\downarrow$ \\ 
		\hline
        \textbf{MadDemon (ours), \textit{winner}}             & 22.65 & 0.48 & \textbf{0.36} & \textbf{2.22 }\\
        IPCV\_IITM                  & 25.15 & 0.60 & 0.66 &         2.36 \\
        Nam                        & \textbf{25.52} & \textbf{0.63} & 0.65 &         2.46 \\
        CVML                       & 24.59 & 0.62 & 0.61 &         2.47 \\
        ACVLab-NPUST               & 24.77 & 0.61 & 0.60 &         2.49 \\
        SeeCout                    & 25.30 & 0.61 & 0.74 &         2.50 \\
        Image Specific NN for RWSR & 24.31 & 0.60 & 0.69 &         2.56 \\
		\hline
        \textbf{MadDemon (ours), \textit{winner}}            & 20.72 & 0.52 & 0.40 & \textbf{2.34} \\
        SeeCout                    & 21.76 & 0.61 & \textbf{0.38} &         2.43 \\
        IPCV\_IITM                  & \textbf{22.37} & \textbf{0.62} & 0.59 &         2.51 \\
        Image Specific NN for RWSR & 21.97 & \textbf{0.62} & 0.61 &         2.59 \\
	\end{tabular}
    \caption{This table reports the quantitative results from the AIM 2019 Challenge on Real World SR \cite{AIM2019RWSRchallenge}. The arrows indicate if high $\uparrow$ or low $\downarrow$ values are desired.}
	\label{tab:aim_results}
	\vspace{-0.2cm}
\end{table}

\subsection{Ablation Study}
\label{sec:ablation_study}
\begin{figure}[!t]
    \centering  
    \begin{sideways}
        \tiny \hspace{-0.37\linewidth} \textit{JPEG} - TDSR \hspace{0.12\linewidth} \textit{JPEG} - SDSR \hspace{0.1\linewidth} \textit{noise} - TDSR \hspace{0.1\linewidth} \textit{noise} - SDSR
    \end{sideways}
    \begin{minipage}[c]{0.19\linewidth}
	    \includegraphics[width=\linewidth]{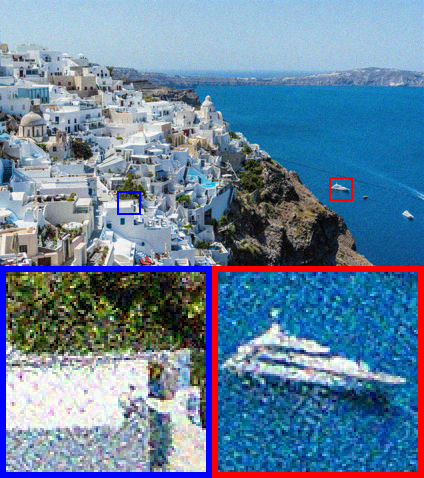}
        \includegraphics[width=\linewidth]{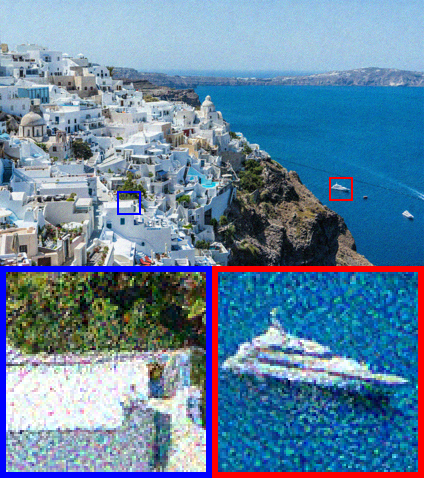}
	    \includegraphics[width=\linewidth]{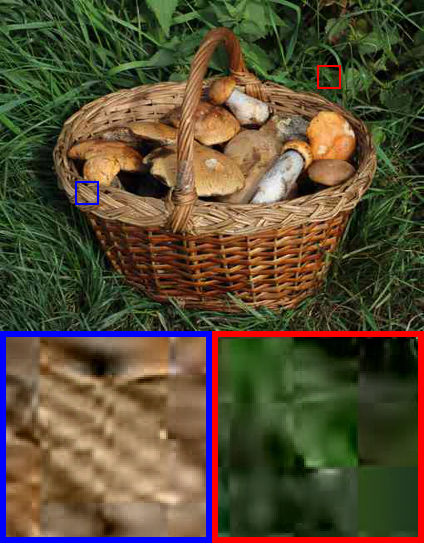}
        \includegraphics[width=\linewidth]{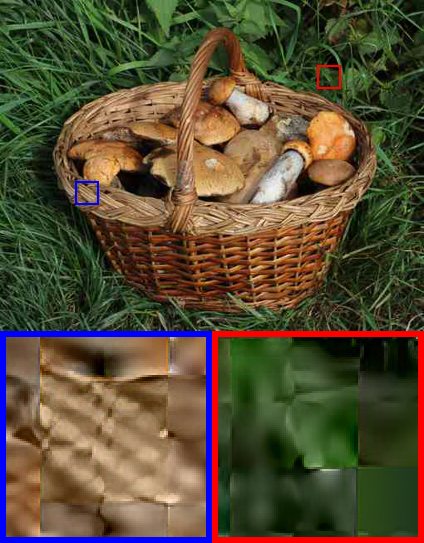}
        \centering \tiny \textit{bicubic} \\ ESRGAN
        \vspace{0.2cm}
	\end{minipage}
 	\hspace{-0.02\linewidth}
 	\begin{minipage}[c]{0.19\linewidth}
        \includegraphics[width=\linewidth]{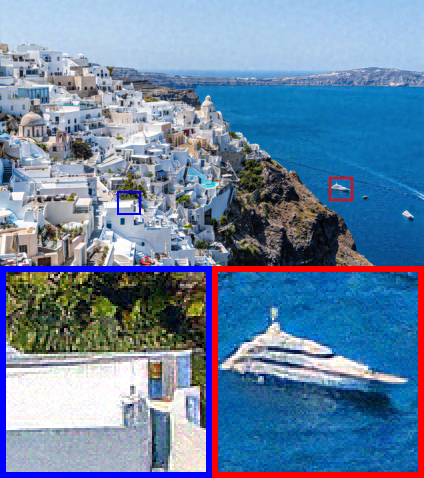}
        \includegraphics[width=\linewidth]{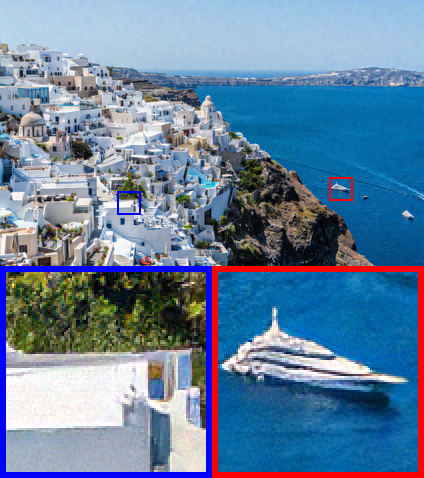}
        \includegraphics[width=\linewidth]{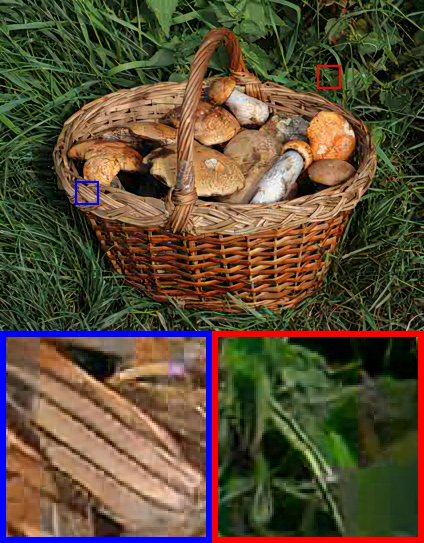}
        \includegraphics[width=\linewidth]{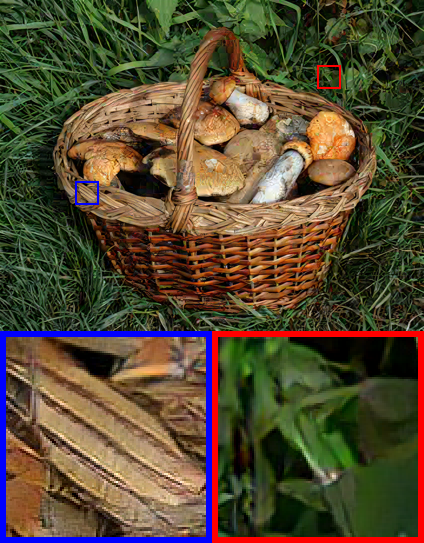}
        \centering \tiny \textit{DSGAN} \\ ESRGAN
        \vspace{0.2cm}
	\end{minipage}
	\hspace{-0.02\linewidth}
 	\begin{minipage}[c]{0.19\linewidth}
        \includegraphics[width=\linewidth]{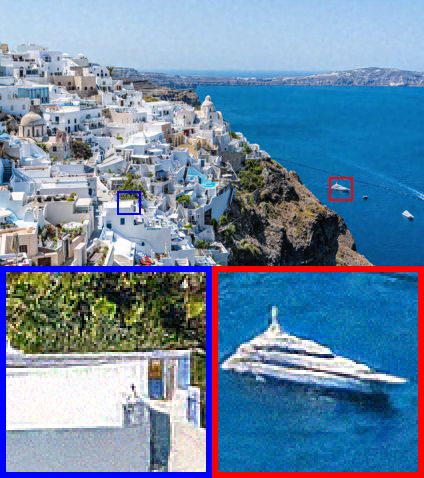}
        \includegraphics[width=\linewidth]{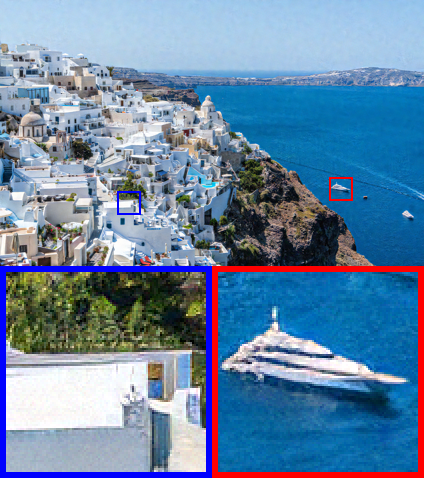}
        \includegraphics[width=\linewidth]{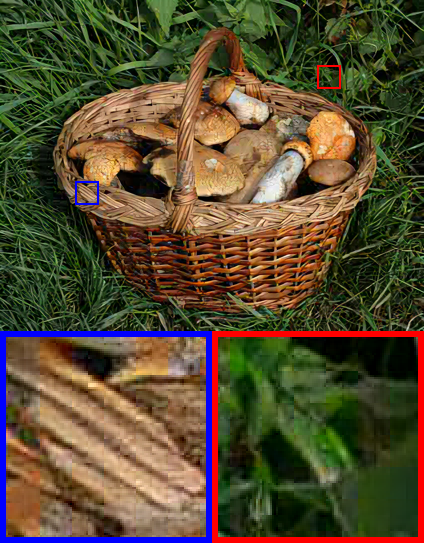}
        \includegraphics[width=\linewidth]{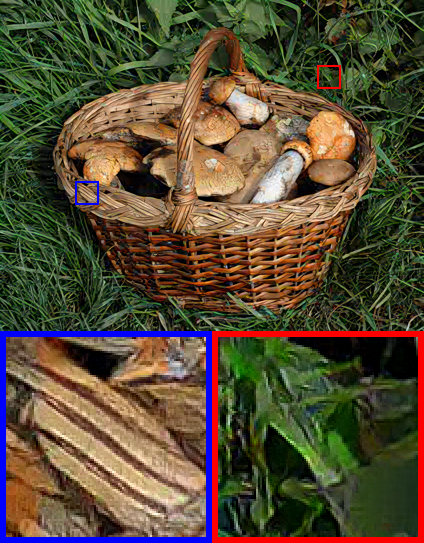}
        \centering \tiny \textit{DSGAN} \\ ESRGAN-FS
        \vspace{0.2cm}
	\end{minipage}
 	\hspace{-0.02\linewidth}
 	\begin{minipage}[c]{0.19\linewidth}
        \includegraphics[width=\linewidth]{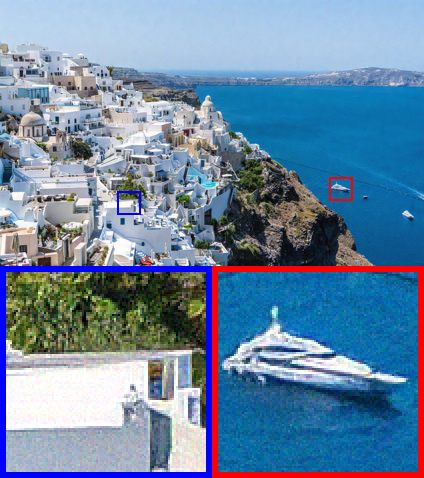}
        \includegraphics[width=\linewidth]{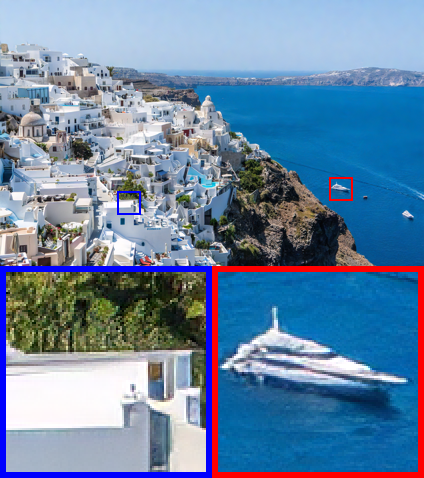}
        \includegraphics[width=\linewidth]{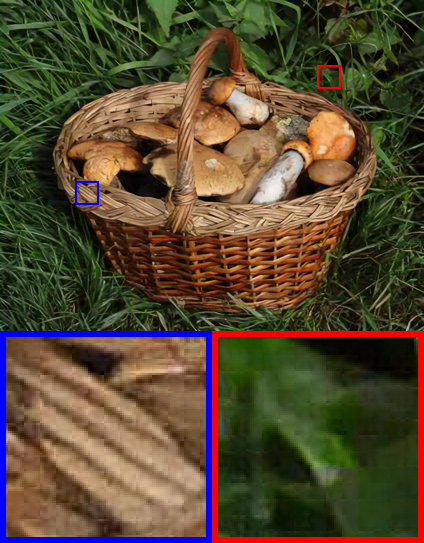}
        \includegraphics[width=\linewidth]{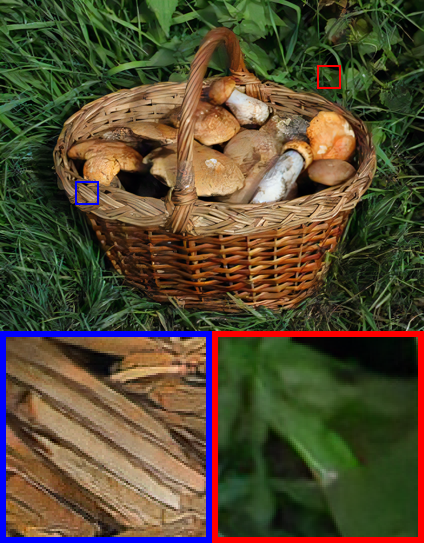}
        \centering \tiny \textit{GT} \\ ESRGAN-FS
        \vspace{0.2cm}
	\end{minipage}
 	\hspace{-0.02\linewidth}
 	\begin{minipage}[c]{0.19\linewidth}
        \includegraphics[width=\linewidth]{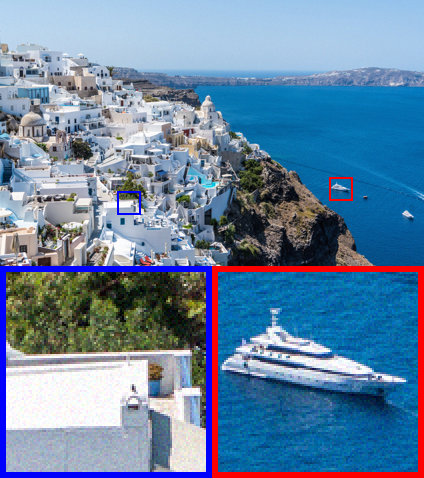}
        \includegraphics[width=\linewidth]{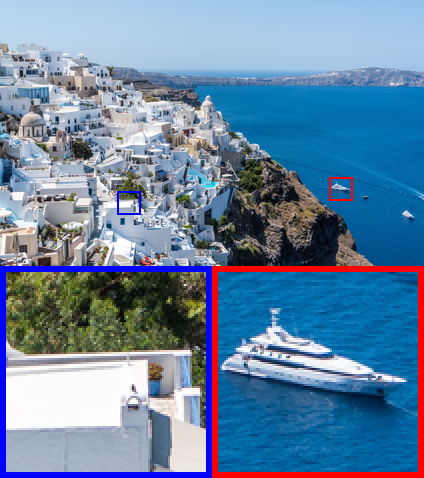}
        \includegraphics[width=\linewidth]{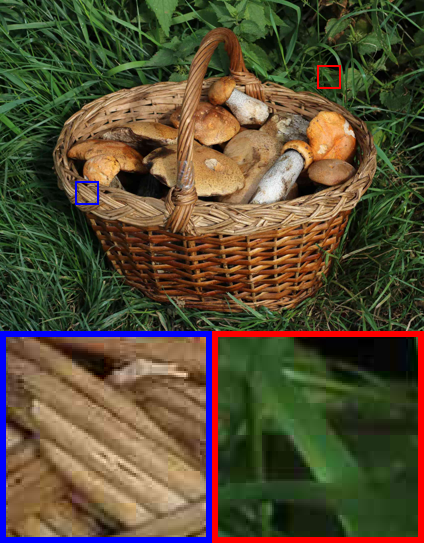}
        \includegraphics[width=\linewidth]{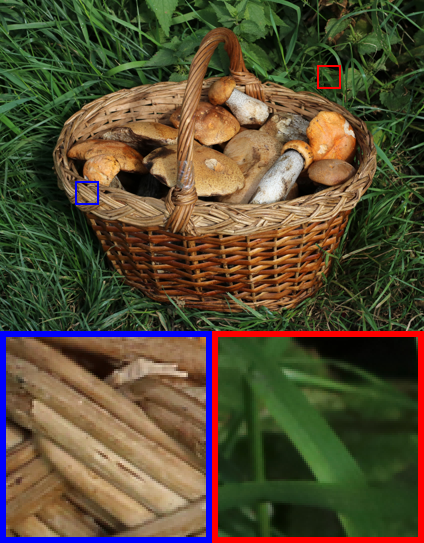}
        \centering \tiny GT \\ ~
        \vspace{0.2cm}
	\end{minipage}
    \caption[Results on DF2K with Sensor Noise]{Ablation study on the DIV2K validation set with sensor noise (\textit{noise}) or compression artifacts (\textit{JPEG}). The first line below the images denotes the \textit{dataset} used for fine-tuning, while the second line refers to the SR~method.}  
    \label{fig:ablation_study_gaussian}    
\end{figure}

In this section, we compare different versions and combinations of our models. We evaluate our method in the same setting as used in Section~\ref{sec:experiments_corrupted}. 
Although we also report PSNR and SSIM, we focus mostly on LPIPS in our discussion as it correlates best with perceptual similarity. As ground truth, we use the corrupted and clean HR images in the SDSR and TDSR settings, respectively. Our quantitative analysis is provided in Table~\ref{tab:ablation_study}. Furthermore, we provide visual results on the DIV2K~\cite{agustsson2017ntire,timofte2017ntire} validation set in Figure~\ref{fig:ablation_study_gaussian} for sensor noise and compression artifacts.

We vary two things in our experiments: the model that is used for SR and the method that is used to generate the HR/LR image pairs. We compare ESRGAN~\cite{WangYWGLDQL18} and ESRGAN-FS (with frequency separation) as our SR models. In each case, one of these models is fine-tuned with one of the following datasets. 

\textit{bicubic}: This standard method results in very poor SR performance. In all cases, it produces strong corruptions in the SR images. This is also reflected in the LPIPS values, which are the worst of all compared methods.

\textit{DSGAN}: In all cases, using DSGAN greatly improves the performance of the method compared to using bicubic downscaling. 
Thereby, ESRGAN-FS tends to produce slightly sharper images than ESRGAN, which leads to a slightly worse LPIPS score for ESRGAN-FS, as the introduced details are often different from ground truth. Furthermore, ESRGAN-FS better matches the source characteristics in the output, which can be seen in the SDSR setting with compression artifacts. ESRGAN does not always manage to produce realistic artifacts, while the artifacts generated by ESRGAN-FS look convincing in all image regions. 

\textit{GT}: In case of input images with sensor noise, this results in images that look very similar to the ones that were generated by using DSGAN. This means that our DSGAN method is very accurate in reproducing the image characteristics of the source images. In case of compression artifacts, the difference is more apparent, as the models trained with the DSGAN dataset introduce some additional corruptions. Interestingly, the models trained with DSGAN produce sharper results than the models trained with GT. This might be because we use bicubic downscaling to generate these image pairs, which not only alters corruptions but also other image characteristics.

\begin{table}[!t]
	\centering
	\begin{sideways}
	\scriptsize \hspace{-0.13\linewidth}TDSR \hspace{0.05\linewidth} SDSR
	\end{sideways}
	\tabcolsep=0.01\linewidth
	\scriptsize
	\begin{tabular}{ll|ccc|ccc}
		\multicolumn{2}{c}{ } & \multicolumn{3}{c}{sensor noise} & \multicolumn{3}{c}{compression artifacts} \\
		dataset & SR method & PSNR$\uparrow$ & SSIM$\uparrow$ & LPIPS$\downarrow$ & PSNR$\uparrow$ & SSIM$\uparrow$ & LPIPS$\downarrow$ \\ 
		\hline
		bicubic & ESRGAN & 18.32 & 0.22 & 0.63 & 23.29 & 0.62 & 0.42 \\
		DSGAN & ESRGAN & 21.36 & 0.38 & 0.23 & 21.88 & 0.57 & 0.35 \\
		DSGAN & ESRGAN-FS & 20.87 & 0.39 & 0.26 & 22.03 & 0.57 & 0.35 \\
		GT & ESRGAN-FS & 22.49 & 0.41 & 0.18 & 23.26 & 0.62 & 0.35 \\
		\hline
		bicubic & ESRGAN & 18.80 & 0.24 & 0.80 & 22.65 & 0.58 & 0.52 \\
		DSGAN & ESRGAN & 22.45 & 0.54 & 0.32 & 21.21 & 0.54 & 0.40 \\
		DSGAN & ESRGAN-FS & 22.52 & 0.52 & 0.33 & 20.39 & 0.50 & 0.42 \\
		GT & ESRGAN-FS & 25.02 & 0.69 & 0.21 & 22.43 & 0.59 & 0.34 \\
	\end{tabular}
    \caption{This table reports the quantitative results of our ablation study. The arrows indicate if high $\uparrow$ or low $\downarrow$ values are desired.}
	\label{tab:ablation_study}
	\vspace{-0.3cm}
\end{table}

\section{Conclusion}
\label{sec:conclusion}

We propose DSGAN to generate paired HR and LR images with similar characteristics. We argue that the relevant characteristics mainly appear in the high frequencies of an image, which we exploit by applying the adversarial loss only on these frequencies. Furthermore, we also apply our idea of frequency separation to the SR model, which allows it to match the target distribution more closely. Our multiple experiments with artificial and natural corruptions demonstrate the effectiveness of our approach for real-world SR. We not only beat the state-of-the-art methods in these experiments, but also won the AIM 2019 Challenge on Real World Super-Resolution~\cite{AIM2019RWSRchallenge}.
\\
\noindent\textbf{Acknowledgments. } This work was partly supported by ETH General Fund and by Nvidia through a GPU grant.
{\small
\bibliographystyle{ieee_fullname}
\bibliography{egbib}
}

\end{document}